%% file: main.tex
\documentclass[runningheads]{llncs}
\usepackage{makeidx}
\usepackage{graphicx}
\usepackage{amsmath,amssymb} 
\usepackage{color}

\usepackage{bbm}
\usepackage{subfigure}
\usepackage{booktabs}
\input{mathmacro}
\usepackage{multirow}
\usepackage{xr}
\usepackage{wrapfig}
\usepackage{enumitem}
\usepackage[misc]{ifsym}
\usepackage{hyperref}
\usepackage{afterpage}

\newcommand{\etal}{\textit{et al.}}
\newcommand{\eg}{\textit{e.g.}}

\begin{document}
	\pagestyle{headings}
	\mainmatter
	\title{Observer Dependent Lossy Image Compression}
	
	\titlerunning{Observer Dependent Lossy Image Compression}
	\authorrunning{M. Weber et al.}
	\author{Maurice Weber$^{1,\,\text{(\Letter)}}$, Cedric Renggli$^1$, Helmut Grabner$^2$ and Ce Zhang$^1$}
	\institute{$^1$Department of Computer Science, ETH Zürich, Switzerland\\$^2$ZHAW School of Engineering, Switzerland\\\email{maurice.weber@inf.ethz.ch}}

	\maketitle

	\begin{abstract}
		Deep neural networks have recently advanced the state-of-the-art in image compression and surpassed many traditional compression algorithms. 
        The training of such networks involves carefully trading off entropy of the latent representation against reconstruction quality. The term quality crucially depends on the observer of the images which, in the vast majority of literature, is assumed to be human. 
        In this paper, we aim to go beyond this notion of compression quality and look at human visual perception and image classification \emph{simultaneously}.
        To that end, we use a family of loss functions that allows to optimize deep image compression depending on the observer and to interpolate between human perceived visual quality and classification accuracy, enabling a more unified view on image compression.
        Our extensive experiments show that using perceptual loss functions to train a compression system preserves classification accuracy much better than traditional codecs such as BPG without requiring retraining of classifiers on compressed images. 
        For example, compressing ImageNet to 0.25 bpp reduces Inception-ResNet classification accuracy by only 2\%.
        At the same time, when using a human friendly loss function, the same compression system achieves competitive performance in terms of MS-SSIM.
        By combining these two objective functions, we show that there is a pronounced trade-off in compression quality between the human visual system and classification accuracy.
	\end{abstract}

	\section{Introduction}
	Image compression algorithms aim at finding representations of images that use as little storage --- measured in bits --- as possible. Opposed to lossless image compression, where the goal is to achieve a high compression rate while requiring perfect reconstruction, lossy image compression enables even higher compression rates by allowing for a loss in reconstruction quality. Recently, image compression based on deep neural networks (DNNs) has achieved remarkable results in both lossless~\cite{mentzer2019} and lossy image compression~\cite{agustsson2017,balle2017,mentzer2018,rippel2017,theis2017,toderici2017}, outperforming many traditional codecs. One distinct advantage of such methods is their flexibility with regards to the term \emph{reconstruction quality} which crucially depends on the observer of the compressed images. Previous research in lossy image compression expressed quality largely in terms of human visual perception and optimized for the human visual system (HVS), using distortion measures such as multiscale structural similarity~\cite{wang2003} (MS-SSIM) or mean squared error (MSE) as training objectives. 
    However, due to recent advances in computer vision systems, increasingly more images are observed solely by machines and bypass humans.
    Consequently, a natural question that arises is whether or not there exists a relation between quality perceived by humans and quality perceived by computer vision systems, and if so, how can we trade off quality between different types of observers? 
    In other words, is a compression system optimized for the human observer also optimal for machines? We investigate these questions by specifically looking at classification of natural images as one of the most well studied tasks in computer vision. 
    The training of modern classifiers is typically a costly and time-consuming undertaking and parameters of the best performing classifiers are often publicly available.
    With that in mind, we are interested in a compression system that generalizes well in the following sense.
    \emph{Firstly}, we want to compress images such that no retraining of classifiers on compressed images is required. 
    \emph{Secondly}, the compression system should be agnostic to classifier architectures. 
    \emph{Thirdly}, it should also generalize well to other visual tasks such as fine-grained visual categorization of natural images. 
    Together, these generalization requirements encourage using publicly available, pretrained classifiers on compressed images from the same domain or on related tasks where classifiers were obtained with transfer learning.
    \begin{figure}[!t]
        \centering
        \subfigure{\includegraphics[width=\linewidth]{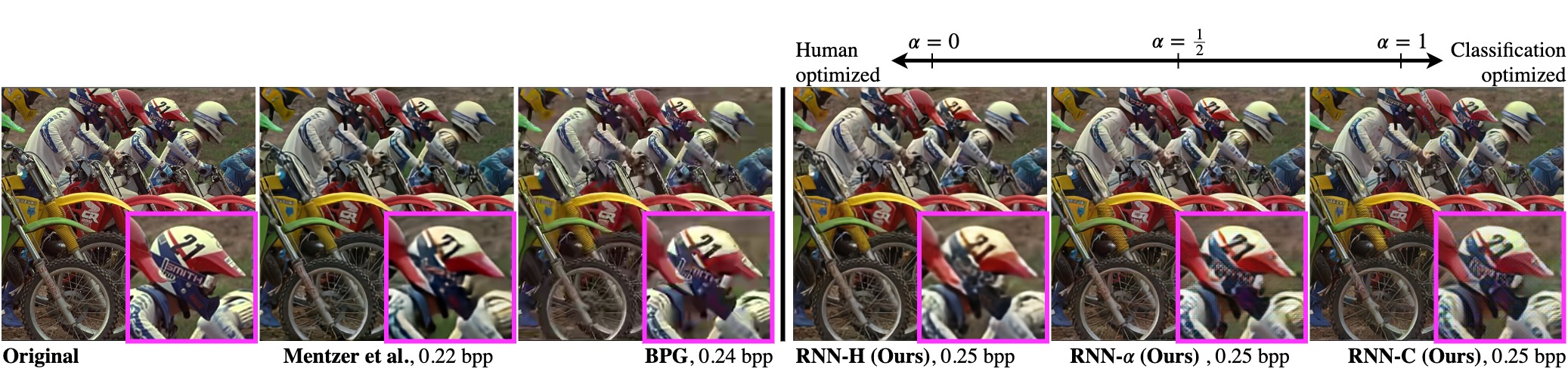}}
        \begin{subfigure}
            \centering
            \begin{minipage}{\linewidth}
                \resizebox{\textwidth}{!}{
                \begin{tabular*}{\linewidth}{@{\extracolsep{\fill}}l c c c c}
                    \toprule
                    & BPG~\cite{bpg} & Mentzer \etal~\cite{mentzer2018} & Ours (for Human) & Ours (for Classification)\\
                    \midrule
                    MS-SSIM & 0.942 & \textbf{0.959} & 0.954 & 0.891 \\
                    Accuracy\footnote{Accuracy uncompressed: 0.803} & 0.707 & 0.719 & 0.715 & \textbf{0.780} \\
                    \bottomrule
                \end{tabular*}
                }
            \end{minipage}
            \label{table:teaser_tradeoff}
        \end{subfigure}
        \caption{Accuracy evaluated on ImageNet-1K with off-the-shelf Inception-ResNet-V2. MS-SSIM on Kodak. Both datasets are compressed at $\sim$0.25 bpp with different methods. Our Classification optimized system induces very low loss in classification accuracy at high compression rates, compared to human optimized approaches.}
    \end{figure}
    Our method for classification oriented compression relies on a feature reconstruction loss using deep features extracted from the hidden layers of a convolutional neural network trained for image classification.
    This type of loss function has been used in the context of super-resolution~\cite{bruna2016,johnson2016,ledig2017}, style-transfer~\cite{gatys2016,johnson2016} and variational autoencoders~\cite{dosovitskiy2016} with remarkable success.
    In order to optimize for human visual perception, we make use of MS-SSIM as a measure of quality perceived by humans, since it has been reported to correlate better with the HVS than MSE. 
    Finally, the convex combination of the two objectives allows to investigate the trade-off between between human visual perception and classification in the context of image compression. In summary, the contributions of our work are threefold:
    \begin{itemize}[leftmargin=*]
        \item We show that training deep image compression with a perceptual loss function preserves classification accuracy much better than human optimized compression systems. In addition, our experiments show that \textit{(1) we do not have to retrain classifiers on compressed images in order to preserve accuracy on highly compressed images}, and \textit{(2) using VGG-based feature reconstruction loss generalizes to other models, indicating that deep CNN features are shared between CNN architectures}.
        \item By looking at the convex combination between human and classification-friendly loss, we present a \textit{simple way to trade off compression quality} in terms of human perception against image classification. Since we only rely on the training objective, our method can be integrated to any learned lossy image compression system.
        \item Our extensive experimental study indicates that \textit{there exists a pronounced trade-off} between compression quality perceived by the human observer and classification accuracy. We show how improved compression quality for the human observer comes at the cost of degraded classification accuracy, and vice versa.
    \end{itemize}
    We emphasize that the contribution of this work is not presenting a new type of loss function nor in a new deep compression architecture. Rather, we make us of existing techniques in order to present a method to trading off compression quality depending on the observer and to show that it is possible to explicitly optimize compression for subsequent classification.

    \section{Related work}
    
    \paragraph{Deep image compression.} Image compression using DNNs has recently become an active area of research. The most popular types of architectures used for image compression are based on autoencoders~\cite{agustsson2017,balle2017,mentzer2018,rippel2017,theis2017} and recurrent neural networks~\cite{johnston2018,toderici2016,toderici2017} (RNNs). Typically, the networks are trained in an end-to-end manner to minimize a pixel-wise notion of distortion such as MSE, MS-SSIM or \(L_1\)-distance between original and decoded image.
    \paragraph{Compression for computer vision.} Image compression in combination with other computer vision tasks has been studied in a number of recent works. Liu \etal~\cite{liu2018} propose an image compression framework based on JPEG that is favorable to DNN classifiers. Also starting from an engineered codec, Liu \etal~\cite{liu2019} propose a 3D image compression framework based on JPEG2000 which is tailored to segmentation of 3-D medical images. Both works differ from ours in that we look at learned image compression, rather than modifying an engineered one. A few examples exist in the literature, where a classifier is learned from features extracted from the encoded representations. Gueguen \etal~\cite{gueguen2018} train a modified ResNet-50 directly on the blockwise discrete cosine transform coefficients from the middle of the JPEG codec. Torfason \etal~\cite{torfason2018} make use of the compressive autoencoder proposed in~\cite{theis2017} and train neural networks for classification and segmentation on the latent (quantized) representations and on the decoded images. These works stand orthogonal to ours in that we do not allow training on compressed versions of images. Rather, we train the compression algorithm such that it maintains information relevant for subsequent classification, keeping the classifiers fixed. We furthermore focus on agnosticity to architectures of inference algorithms. Finally, since compression artifacts typically compromise the performance of classifiers, Dodge and Karam~\cite{dodge2016} study the effect of JPEG compression on image classification with neural networks.
    \paragraph{Feature reconstruction loss.} This class of similarity functions makes use of deep features extracted from convolutional neural networks. Recent advances in generative modelling have shown that using this type of loss functions, high quality images can be generated and have been applied to a variety of tasks. Gatys \etal~\cite{gatys2015,gatys2016} apply the idea to style transfer and texture synthesis, while Johnson \etal~\cite{johnson2016} and Bruna \etal~\cite{bruna2016} achieve remarkable results in super resolution~\cite{johnson2016,bruna2016} and style transfer~\cite{johnson2016}. Ledig \etal~\cite{ledig2017} further develop the idea and enhance the CNN feature loss with adversarial training to achieve state-of-the-art results in single image super resolution. In the image compression domain, steps in this direction have also been made. Agustsson \etal~\cite{agustsson2018}, Santurkar \etal~\cite{santurkar2018} and Liu \etal~\cite{liu2018_losses} enhance pixel-wise distortion and adversarial training with a feature reconstruction loss. Furthermore, Chinen \etal~\cite{chinen2018} and Zhang \etal~\cite{zhang2018} both propose a similarity metric based on deep features extracted from VGG-16 trained for image classification. These works have in common that their focus is on the human observer, while we exploit properties of feature reconstruction loss in the context of compression geared towards subsequent image classification. Feature reconstructions loss has also been used in the context of compression artifact removal. Galteri \etal~\cite{galteri2017} train a generative adversarial network in combination with a VGG-based perceptual loss function to remove compression artifacts in images. It is shown that this can significantly increase the quality of compressed images in terms of MS-SSIM and in terms of object detection accuracy. However, contrary to our work, no clear trade-off between the human observer and image classification is investigated.
    
    \section{Method}
    In this section, we outline our approach to compressing images for human visual per\-ception, classification accuracy and the interpolation between the two. Through\-out this paper we adopt the compression architecture proposed by Tode\-rici \etal~\cite{toderici2017}, based on recurrent neural networks. We emphasize that we only focus on the objective functions to account for different types of observers.
    \paragraph{Compression framework.}
    Let \(\gX \subseteq \bR^d\) denote a set of training images, \(\gZ\subseteq\bZ\) the quantization levels and \(d\colon\bR^d \times\bR^d\to\bR\) a notion of distortion between images. Our goal is to find a compression system consisting of an encoder \(E\colon\bR^d\to\bR^m\) that maps input images \(\mathbf{x}\) to their latent representation \(\mathbf{z}=E(\mathbf{x})\), a quantizer \(q\colon\bR^m\to\gZ^m\) that discretizes \(\mathbf{z}\) to \(\mathbf{\hat{z}}=q(\mathbf{z})\), and a decoder \(D\colon\gZ^m\to\bR^d\) that maps the quantized representation back to image space, \(\mathbf{\hat{x}} = D(\mathbf{\hat{z}})\). The goal is then to minimize the rate-distortion trade-off over the training set \(\gX\), i.e. for \(\beta\geq0\), we want to minimize \(\sum_{\mathbf{x} \in \gX} d(\mathbf{x},\,\mathbf{\hat{x}}) + \beta\,H(\mathbf{\hat{z}})\), where \(H\) denotes the entropy. 
    As a compression architecture, we adopt the RNN-based model proposed in~\cite{toderici2017} with gated recurrent units (GRUs), allowing for variable bitrates.
    An input image $\mathbf{x}$ is passed through the encoder and quantizer, mapping the latent codes stochastically to $\gZ^m=\{-1,+1\}^{m}$. The quantized representation is subsequently decoded, yielding an estimate of the original image.
    This is repeated with the residual error fed to the encoder to obtain an estimate at the next bitrate, using information from the hidden states of the previous iterations. Formally, a single iteration at unrolling step $t\geq 1$, can be represented as $\mathbf{\hat{x}_t}=\mathbf{\hat{x}_{t-1}} + D_t(Q(E_t(\mathbf{r}_t)))$ with $\mathbf{\hat{x}_0}=\mathbf{0}$ and $\mathbf{r_1}=\mathbf{x}$ and where $E_t$ and $D_t$ are encoder and decoder carrying information from the previous unrolling steps.
    Finally, we remark that, since $\gZ$ contains a finite number of quantization levels, we set $\beta=0$ in the training objective.
    \paragraph{Optimizing for human visual perception.}
    In order to optimize the compression system for the human observer, we choose a measure of distortion that approximately models human visual perception. The multiscale structural similarity index (MS-SSIM)~\cite{wang2003} is based on the assumption that the human eye is adapted for extracting structural information from images and incorporates image details at multiple resolutions. It is furthermore reported to correlate better with human visual perception than MSE. Since MS-SSIM is differentiable, we follow \cite{johnston2018,mentzer2018,rippel2017} and minimize directly $d_H(\mathbf{x},\,\mathbf{\hat{x}}) = 1 - \text{MS-SSIM}(\mathbf{x},\,\mathbf{\hat{x}})$.
    We refer to compression optimized with \(d_H\) as \textbf{RNN-H}.
    An alternative approach would be to use other, human-centric distortion metrics such as LPIPS~\cite{zhang2018} or the approach proposed in~\cite{chinen2018}. However, as these approaches are based on CNNs they bear the additional challenge of dealing with checkerboard-like artifacts~\cite{odena2016}.
    
    \paragraph{Optimizing for classification.}
    Suppose we are given a CNN classifier \(f\) trained on a set of images and labels \((\gX', \gY')\) and corresponding training and validation splits \((\gX'_{train}, \gY'_{train})\) and \((\gX'_{val}, \gY'_{val})\). When we optimize compression for classification accuracy, we are interested in finding an encoder, quantizer and decoder such that the accuracy evaluated on the decoded validation set \(D(q(E(\gX'_{val})))\) is maintained as well as possible, \emph{without} further retraining the classifier on decoded images. Formally, we wish to maximize $\sum_{\mathbf{x}\in\gX'_{val}}\mathbbm{1}\{f(\mathbf{x}) = f(\mathbf{\hat{x}})\}$.
    We are thus not interested in matching decoded and original images on a pixel-wise basis, but rather on preserving features which are relevant for subsequent classification. Image classification is a task which is typically invariant to translations and local deformations (see \eg~\cite{bruna2013,mallat2016}), which motivates the use of an objective function with similar properties. For example, using a pixel-wise distortion, such as MSE, which is not invariant to such deformations would be a suboptimal choice. Furthermore, minimizing MSE encourages the generator to produce images that are pixel-wise averages of plausible solutions \cite{dosovitskiy2016}, resulting in overly smooth images. In other words, high frequency information such as textures  will tend to get lost in the compression process. While this is less problematic for the HVS, which is more susceptible to low frequency changes, CNNs are sensitive to any change in frequency~\cite{liu2018}.\par
    Features learned by convolutional neural networks~\cite{lecun2010} for image classification provide a promising alternative. The intuition is that, if such features are maintained in the compression process, then the compressed representations are encouraged to encode information relevant to classification rather than to the human observer. Moreover, it is known that CNNs provide stability to small geometric deformations and translations, thanks to rectification and pooling units~\cite{bruna2013}. This is beneficial for our purpose, since we do not want to put too much emphasis on such deformations as they do not affect classification. Finally, feature reconstruction loss typically leads to high frequency artifacts (\cite{dosovitskiy2016} and references therein) and checkerboard patterns \cite{odena2016}. While this harms human perceived visual quality, our experiments indicate that this is not the case for classification. These considerations make distortion measures based on CNN features promising candidates for classification oriented image compression.\par
    In order to define a distortion measure that incorporates these properties, we fix a CNN classifier \(f_L\) trained on a dataset \((\gX'',\,\gY'')\). Denote by \(\phi_{i}\) the responses of the i-th convolutional layer after activation and let \(\gI\) be a set of such layers. Note that \(\gI\) is not required to include all layers. We then define the distortion measure associated with the loss network \(f_L\) and layers \(\gI\) to be MSE in feature space
    \begin{equation}
        d_{C,\,\gI}(\mathbf{x},\,\mathbf{\hat{x}}) = \sum_{i \in \gI} \gamma_i \|\phi_i(\mathbf{x}) - \phi_i(\mathbf{\hat{x}})\|_2^2,
        \label{eq:feature_reconstruction_loss}
    \end{equation}
    where \(\gamma_i:=(H_i \times W_i \times C_i)^{-1}\) and \(H_i,\,W_i,\,C_i\) represent the spatial dimensions of the corresponding layer. Note that we do not restrict the loss network to be trained on the same dataset as the compression system or the classifier \(f\), however we do require that \(\gX'' \cap \gX'_{val} = \varnothing\). Furthermore, the classifier \(f\) might have a different underlying architecture than the loss network \(f_L\). This formulation allows to investigate the generalizability of the compression system to new datasets and CNN architectures. We refer to compression optimized with \(d_{C,\,\gI}\) as \textbf{RNN-C}.
    \paragraph{From human visual perception to classification.}
    \label{sec:interpolation_loss}
    In a scenario where images are consumed by both humans and classifiers, we would like to be able to trade off reconstruction quality between the two observers. In other words, we want to have a compressed representation of an image that contains features relevant for classification \emph{and} looks visually pleasing for the human observer. At the same time, this enables us to investigate the relation between human visual perception and classification accuracy. For that purpose, we consider the convex combination between distortions \(d_H\) and \(d_{C,\,\gI}\)
    \begin{equation}
        d_{\alpha,\,\gI}(\mathbf{x},\,\mathbf{\hat{x}}) = (1-\alpha) \cdot \lambda_H \cdot d_H(\mathbf{x},\,\mathbf{\hat{x}}) + \alpha \cdot d_{C,\,\gI}(\mathbf{x},\,\mathbf{\hat{x}})
        \label{eq:interpolation_loss}
    \end{equation}
    and control the trade-off with the parameter \(\alpha \in [0,\,1]\). The parameter \(\lambda_H\) is a scaling parameter which keeps the two losses on the same magnitude and is set to 5,000. We refer to compression optimized with \(d_{\alpha,\,\gI}\) as \textbf{RNN-}\(\boldsymbol{\alpha}\).
    
    \section{Experiments}
    \label{sec:experiments}
    \begin{figure}[!t]
        \centering
        \includegraphics[width=\linewidth]{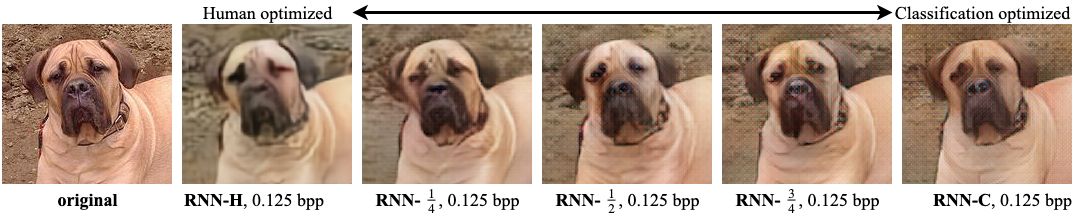}
        \caption{Sample image from the Stanford Dogs dataset. RNN-H results in smoother and blurrier images, RNN-C on the other hand produces sharp images but suffers from checkerboard-like artifacts stemming from the CNN based loss function.}
        \label{fig:illustration1}
    \end{figure}
    In this section we experimentally validate our approach to trading off compression quality between human visual perception and classification accuracy, making use of the proposed family of loss functions. All models are implemented in Python using the Tensorflow~\cite{tensorflow2015} library.\footnote{The source code is available at~\url{https://github.com/DS3Lab/odlc}.}
    \paragraph{Image compression.}We use the RNN compression architecture proposed by Toderici \etal~\cite{toderici2017} with GRUs and the additive reconstruction framework. Our implementation differs from the original version in two aspects. Firstly, during  training, we feed as input the full resolution images, rather than 32\(\times\)32 image patches. And secondly, instead of optimizing the \(L_1\)-distance in image space, we use the family of loss functions~(\ref{eq:interpolation_loss}) as training objective. Furthermore, we do not use the lossless entropy coding scheme proposed in their original work.  While this would likely result in reduced bitrates, and thereby further improve our results, we omit this in order to reduce complexity and focus exclusively on the distortion during training. If not stated otherwise, we train the networks for 8 unrolling steps, yielding rates between 0.125 and 1.0 bpp. As training data \(\gX\), we use the training split of the ILSVRC-2012~\cite{russakovsky2015} dataset, commonly known as ImageNet-1K. We preprocess the images by resizing such that the smallest side equals 256 pixels and aspects are preserved using bilinear interpolation. During training, we take random crops of size 224\(\times\)224 and randomly flip them horizontally. During validation, we use the central crop of size 224\(\times\)224. We follow~\cite{mentzer2018} and normalize with a mean and variance obtained from a subset of the training set. We train all our networks using the Adam optimizer~\cite{kingma2014} for three epochs with the learning rate set to 4e-4 and minibatches of size four. All models are trained on eight Nvidia Titan X GPUs with 12GB RAM.
    \paragraph{Measures of distortion.} We train the compression networks using the loss function defined in equation~(\ref{eq:interpolation_loss}) and use VGG-16 trained on the ImageNet-1K training split as our loss network \(f_L\). Preliminary experiments have shown that choosing \(\gI=\{\phi_{1.1},\,\phi_{5.1}\}\) in~(\ref{eq:feature_reconstruction_loss}), where \(\phi_{i.j}\) denotes the j-th convolutional layer after activation in the i-th block of VGG-16, performed best. Including the entire set of convolutions or only the first or last layer did not yield any improvements. We provide detailed plots to compare the different loss compositions with respect to accuracy in the supplementary materials. The weights of the loss network are frozen and left unchanged during training. We experiment with different values for the parameter \(\alpha\), starting the training each time from scratch. Namely, in order to optimize for human visual perception, we set \(\alpha=0\), while for classification oriented compression, we set \(\alpha=1\). To investigate the trade-off between human vision and classification, we train with \(\alpha \in \{\frac{1}{4},\,\frac{1}{2},\,\frac{3}{4}\}\), also starting training from scratch each time. 
    \paragraph{Comparison with other methods.} We compare our approach to the traditional compression algorithms JPEG~\cite{wallace1992}, WebP~\cite{webp} and BPG~\cite{bpg} which achieves state-of-the-art performance in HVS oriented compression. Following~\cite{mentzer2018,rippel2017}, BPG is used in the non-default 4:4:4 chroma format. Finally, we also compare against the state-of-the-art learned compression method presented in Mentzer \etal~\cite{mentzer2018}, using their publicly available weights and code. Since the available weights do not compress images below 0.3 bpp, we train two models with the same hyperparameters but different number of bottleneck channels to achieve lower bitrates.
    \paragraph{Datasets.} We evaluate our approach on four publicly available datasets. For classification we use the ImageNet-1K dataset~\cite{russakovsky2015}, as well as two datasets used for fine-grained categorization, CUB-200-2011~\cite{wah2011} and Stanford Dogs~\cite{Khosla2011}. In order to evaluate the performance in terms of human visual perception we use the Kodak Photo CD dataset~\cite{kodak_photos} and the ImageNet-1K validation split.
    \paragraph{CNN architectures.} On ImageNet-1K, we use DenseNet-121~\cite{huang2017}, Inception-ResNet-V2~\cite{szegedy2017}, Inception-V3~\cite{szegedy2016}, MobileNet-V1~\cite{howard2017}, ResNet-50~\cite{he2016a}, Xception~\cite{chollet2017} and VGG-16~\cite{simonyan2015} for inference and use the weights provided by the Keras Library~\cite{keras}. For fine-grained categorization on CUB-200-2011 and Stanford Dogs, we use Inception-V3, Inception-ResNet-V2, MobileNet-V1, ResNet-50 and VGG-16. To obtain the classifiers, we use ImageNet pre-trained networks and fine-tune all layers on the original uncompressed training split.
    \paragraph{Evaluating classification accuracy.} In order to compare the different compression algorithms with regard to classification accuracy, we evaluate a collection of CNN architectures on datasets compressed with different algorithms and at different bitrates. Note that all classifiers are trained on the uncompressed respective training datasets, \emph{without} retraining on decoded data. The evaluation procedure is as follows. Since generally, the images do not have the same resolution, we resize them such that the smaller side equals \(S_{comp}\) and aspects are preserved. We then take the central crop of size \(S_{comp}\times S_{comp}\) yielding square images. After this step, given a compression algorithm, we encode the images for a predefined grid of quality parameters and compute the bpp values for each image and quality parameter. For each quality level, we subsequently take the average over the entire validation set, yielding the final bpp values. Finally, we decode and take the central crop of size \(S_{inf}\times S_{inf}\) of the decoded image, which is then fed to the classifier. This results in a set of (bpp, accuracy) points for each classifier and compression method. For CNNs that expect inputs of size \(S_{inf}=299\) we set \(S_{comp}=336\) and for those with \(S_{inf}=224\), we set \(S_{comp}=256\).
    \paragraph{Evaluating human visual perception.} The procedure to compare the compression methods for human perceived visual quality is as follows. To account for the variable resolution on ImageNet-1K, we resize each image with bilinear interpolation such that the smallest side equals 256 pixels and aspects are preserved. We then take the central crop of size 256\(\times\)256. Since the Kodak images are all of equal resolution, we skip this first resizing step and keep the original resolution. We then compress the images using a predefined grid of quality parameters and compute their bpp values which are averaged over the validation set. Finally, we compute the MS-SSIM scores between decoded and original (resized) image. This yields a set of (bpp, MS-SSIM) points for each compression method.
    \subsection{Results} We start by investigating the trade-off between human perception and classification accuracy using compression trained with an increasingly more classification friendly loss. We then look at compression in terms of classification accuracy, followed by our results on human perception.
    \paragraph{From human visual perception to classification.}
    \begin{figure}[!t]
        \centering
        \subfigure[Inception-ResNet-V2, \(\sim\)0.25 bpp]{\includegraphics[width=0.49\linewidth]{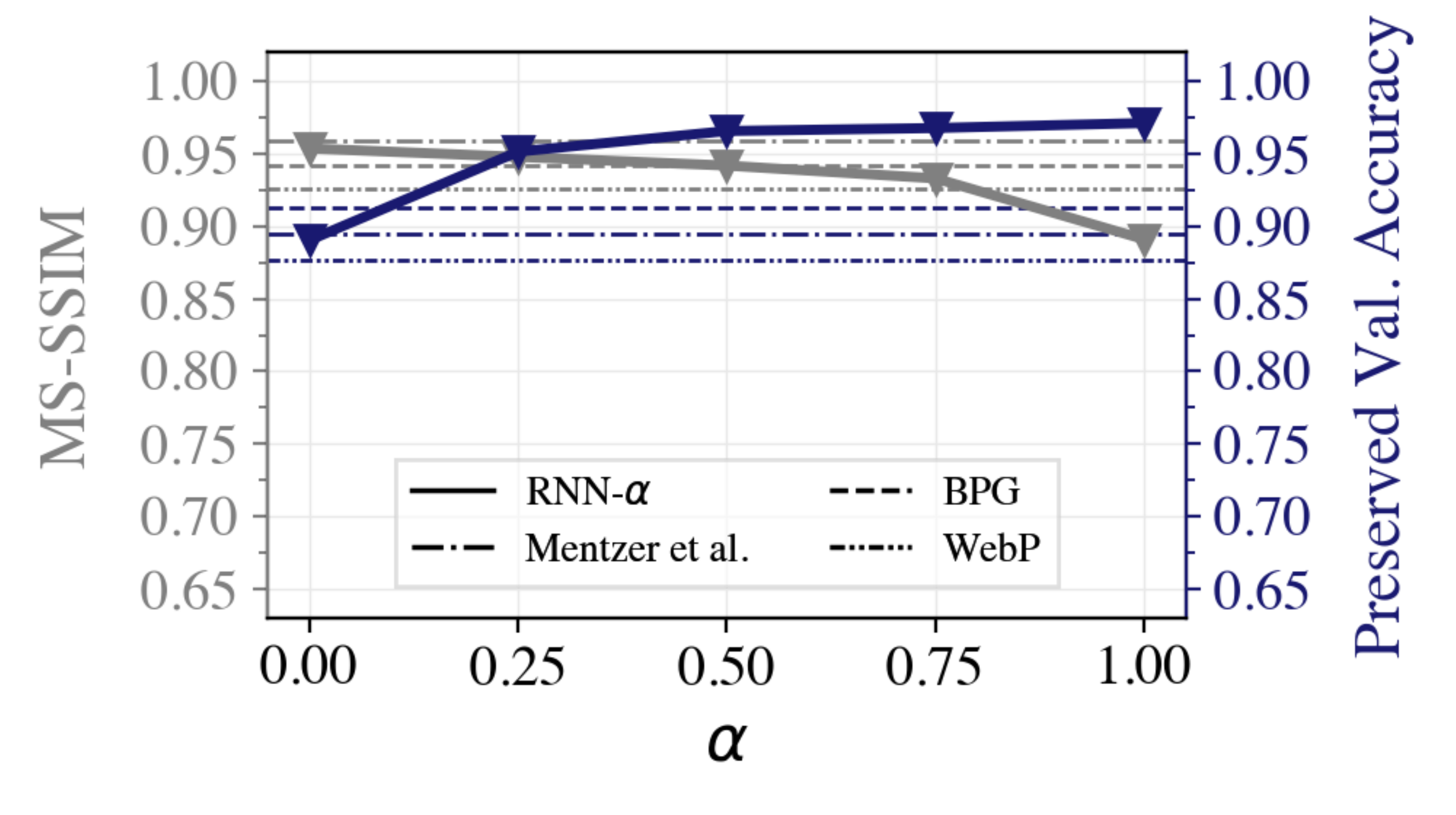}}%
        \hfill
        \subfigure[Inception-ResNet-V2, \(\sim\)1.0 bpp]{\includegraphics[width=0.49\linewidth]{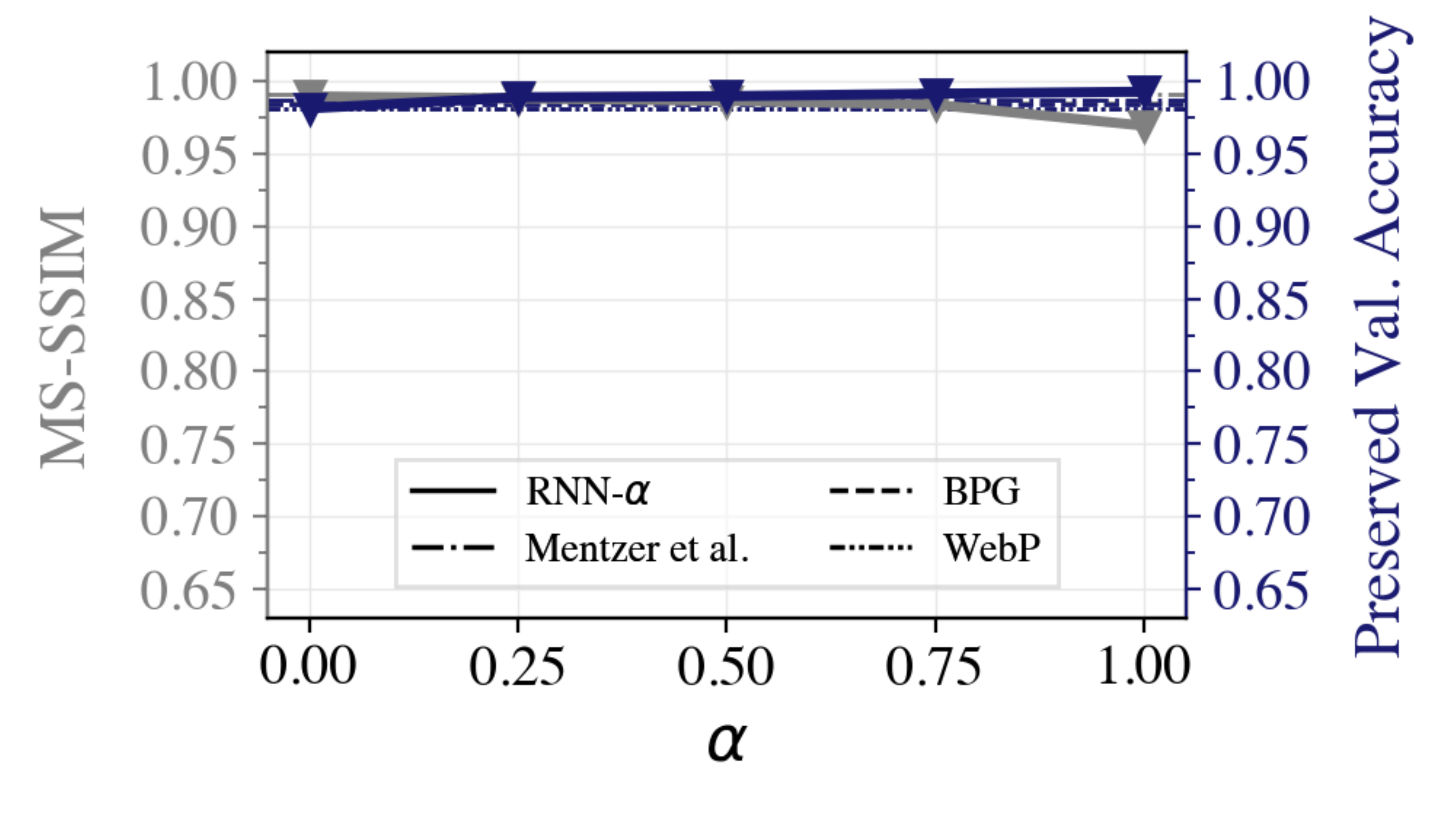}}\\%
        \subfigure[DenseNet-121, \(\sim\)0.25 bpp]{\includegraphics[width=0.49\linewidth]{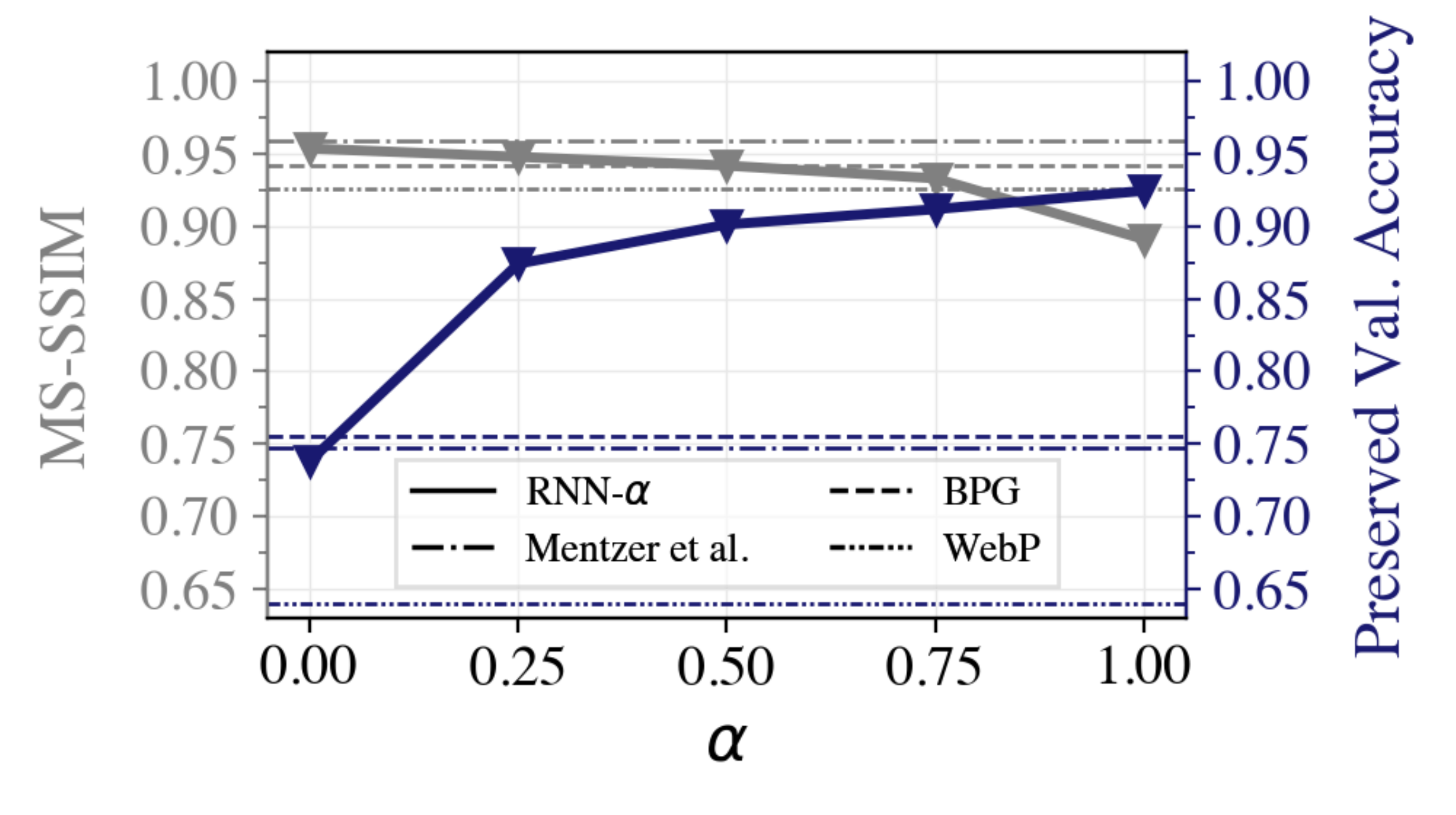}}%
        \hfill
        \subfigure[DenseNet-121, \(\sim\)1.0 bpp]{\includegraphics[width=0.49\linewidth]{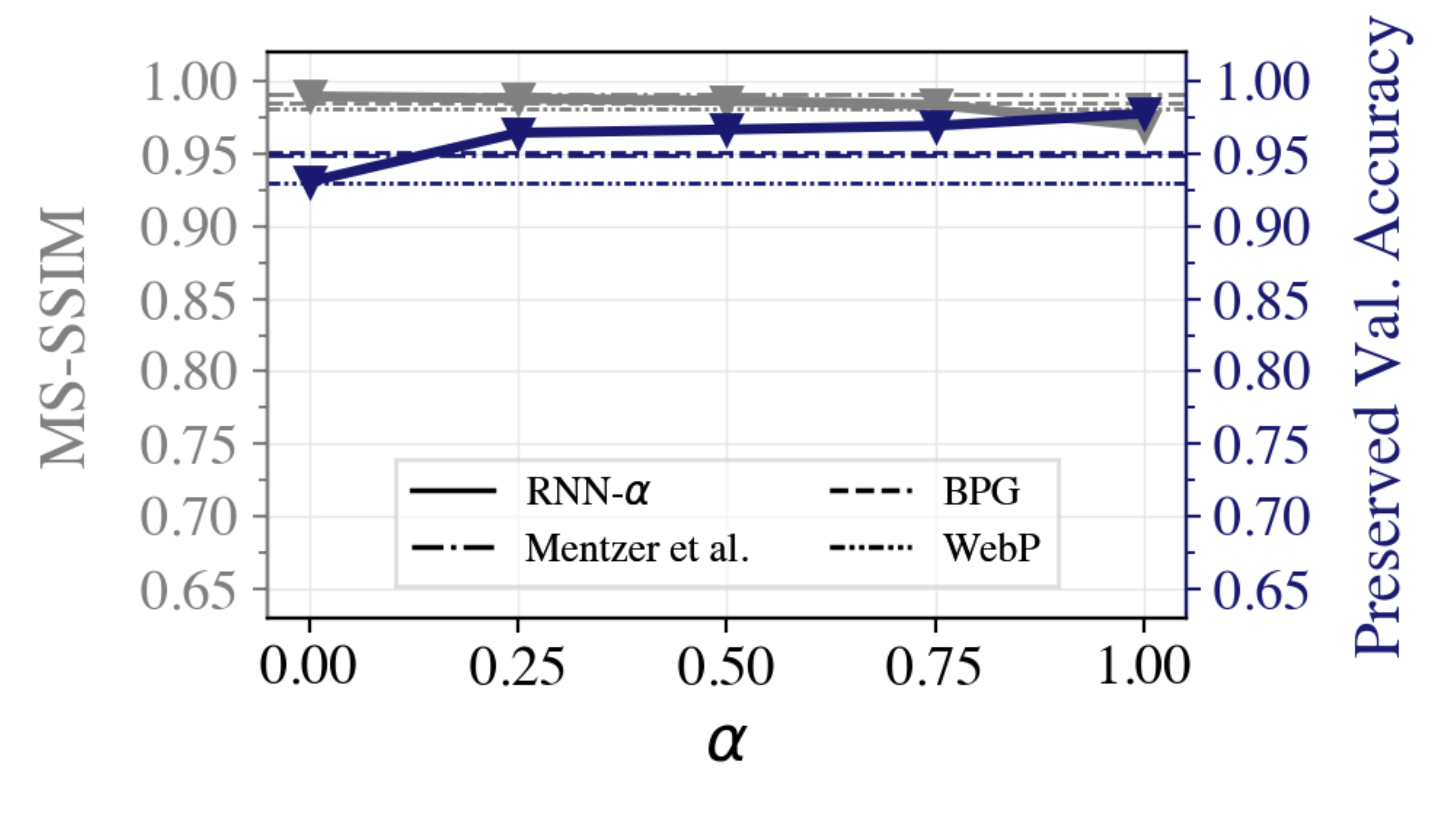}}
        \caption{MS-SSIM evaluated on Kodak (left axis, grey), validation accuracy evaluated on ImageNet-1k (right axis, blue). As $\alpha$ increases, MS-SSIM decreases, while validation accuracy increases. The trade-off is especially pronounced for the low bitrates and DenseNet-121 is in general more sensitive to compression than Inception-ResNet-V2.}
        \label{fig:tradeoff_plots}
    \end{figure}
    In order to investigate the relation between compression quality perceived by humans in terms of MS-SSIM, and by CNN classifiers, we train the compression networks with loss functions that interpolate between human friendly and classification friendly loss, i.e. for values of \(\alpha\) in \(\{0,\,\frac{1}{4},\,\frac{1}{2},\,\frac{3}{4},\,1\}\). This trade-off can be seen qualitatively in Figure~\ref{fig:illustration1}. Optimizing for MS-SSIM, results in images that appear smoother and more blurry. Classification optimized compression on the other hand results in sharper images but suffers from checkerboard-like artifacts. This type of degradation is a known issue for feature visualization and super resolution (see \eg~\cite{odena2016}) and -- in our case -- stems from the convolution based loss function which incurs artifacts in gradients. In order to quantitatively investigate the trade-off, we visualize the relation in Figure~\ref{fig:tradeoff_plots}. We plot MS-SSIM on Kodak (left axis) and ImageNet-1K validation accuracy (right axis) against the tradeoff parameter \(\alpha\) corresponding to RNN compression trained with different loss functions. The Figures indicate that we can indeed trade off accuracy against MS-SSIM by optimizing compression with our family of loss functions. Interestingly, we observe that by increasing the trade-off parameter \(\alpha\) from 0 to 0.25, we substantially increase accuracy while the reduction in MS-SSIM is relatively small. The same holds for the other direction. Finally, we observe that the trade-off is much more pronounced in the low bitrate regime.
    \begin{table}[!t]
        \caption{Validation accuracy on ImageNet-1K. Our RNN-C compression consistently outperforms all other methods across bitrates and architectures.}
        \label{table:accuracy_imagenet}
        \begin{minipage}{\textwidth}
        \centering
        \resizebox{\textwidth}{!}{
        \begin{tabular}{ l  c  c  c  c  c  c  c  c }
            \toprule
            \multicolumn{9}{c}{ImageNet-1K Validation Accuracy}\\
            \toprule
            & \multicolumn{5}{c}{224\(\times\)224 input} & \multicolumn{3}{c}{299\(\times\)299 input}\\
            \cmidrule(lr){2-6} \cmidrule(lr){7-9}
            & bpp & DenseNet-121 & MobileNet & ResNet-50 & VGG-16\footnote{Loss network used to train RNN-C compression.} & bpp & Inception-V3 & Xception\\
            \toprule
            \multicolumn{9}{c}{Low Bitrates (\(\sim\)0.13 bpp)}\\
            \midrule
            RNN-C  & 0.125 & \textbf{0.5914} & \textbf{0.4903} & \textbf{0.6092} & \textbf{0.6009} & 0.125  & \textbf{0.6691}  & \textbf{0.6815}\\
            RNN-H & 0.125 & 0.4109 & 0.3221 & 0.4309 & 0.3797 & 0.125 & 0.5550  & 0.5722\\
            Mentzer \etal~\cite{mentzer2018} & 0.142 & 0.4744 & 0.3774 & 0.4931 & 0.4387 & 0.140 & 0.6069 & 0.6269\\
            BPG & 0.157 & 0.4661 & 0.3421 & 0.4711 & 0.4448 & 0.132 & 0.5750 & 0.6050\\
            JPEG & 0.136 & 0.0480 & 0.0493 & 0.0426 & 0.0320 & 0.113 & 0.2675 & 0.2166\\
            \toprule
            \multicolumn{9}{c}{Medium Bitrates (\(\sim\)0.65 bpp)}\\
            \midrule
            RNN-C & 0.625 & \textbf{0.7252} & \textbf{0.6256} & \textbf{0.7246} & \textbf{0.6998} & 0.625 & \textbf{0.7678} & \textbf{0.7787}\\
            RNN-H & 0.625 & 0.6709 & 0.5688 & 0.6744 & 0.6450 & 0.625 & 0.7434 & 0.7543\\
            Mentzer \etal~\cite{mentzer2018} & 0.652 & 0.6842 & 0.5909 & 0.6975 & 0.6670 & 0.648 & 0.7567 & 0.7652\\
            BPG & 0.725 & 0.6857 & 0.5834 & 0.6894 & 0.6634 & 0.581 & 0.7377 & 0.7523\\
            WebP & 0.589 & 0.6306 & 0.5263 & 0.6323 & 0.6247 & 0.602 & 0.7268 & 0.7429\\
            JPEG & 0.582 & 0.6166 & 0.5111 & 0.6333 & 0.6365 & 0.686 & 0.7390 & 0.7476\\
            \toprule
            \multicolumn{9}{c}{High Bitrates (\(\sim\)1.0 bpp)}\\
            \midrule
            RNN-C & 1.000 & \textbf{0.7303} & \textbf{0.6347} & \textbf{0.7316} & \textbf{0.7037} & 1.000 & \textbf{0.773} & \textbf{0.7840}\\
            RNN-H & 1.000 & 0.6998 & 0.5984 & 0.7018 & 0.6732 & 1.000 & 0.7631 & 0.7728\\
            Mentzer \etal~\cite{mentzer2018} & 1.037 & 0.7076 & 0.6183 & 0.7152 & 0.6841 & 1.034 & 0.7667 & 0.7767\\
            BPG & 1.048 & 0.7085 & 0.6151 & 0.7168 & 0.6841 & 1.066 & 0.7618 & 0.7756\\
            WebP & 0.997 & 0.6930 & 0.6050 & 0.7039 & 0.6829 & 1.055 & 0.7589 & 0.7699\\
            JPEG & 1.087 & 0.6808 & 0.5865 & 0.6963 & 0.6918 & 0.962 & 0.7517 & 0.7622\\
            \toprule
            Original & - & 0.7453 & 0.6590 & 0.7465 & 0.7088 & - & 0.7786 & 0.7907\\
            \bottomrule
        \end{tabular}}
        \end{minipage}
    \end{table}
    \paragraph{ImageNet classification.} 
    Table~\ref{table:accuracy_imagenet} shows the classification accuracies for a wider collection of CNN architectures. We see that our RNN-C outperforms both the traditional codecs BPG, WebP and JPEG as well as our RNN-H and the deep image compression method proposed in \cite{mentzer2018}, across all architectures and bitrates considered. In the case of the loss network VGG-16 this is to be expected, since we explicitly train the compression network to produce images whose VGG-features -- which are fed to the fully connected layers for classification -- match the ones from their uncompressed version. Interestingly however, we see that RNN-C generalizes well to architectures different from the loss network and maintains the accuracy remarkably well, indicating that hidden representations are shared among CNN architectures. It is again noticeable that the advantage of RNN-C is much more pronounced for low bitrates.
    \begin{figure}[!t]
        \centering
        \subfigure[Stanford Dogs Acc.]{\label{fig:accuracy_dogs}\includegraphics[width=.3\linewidth]{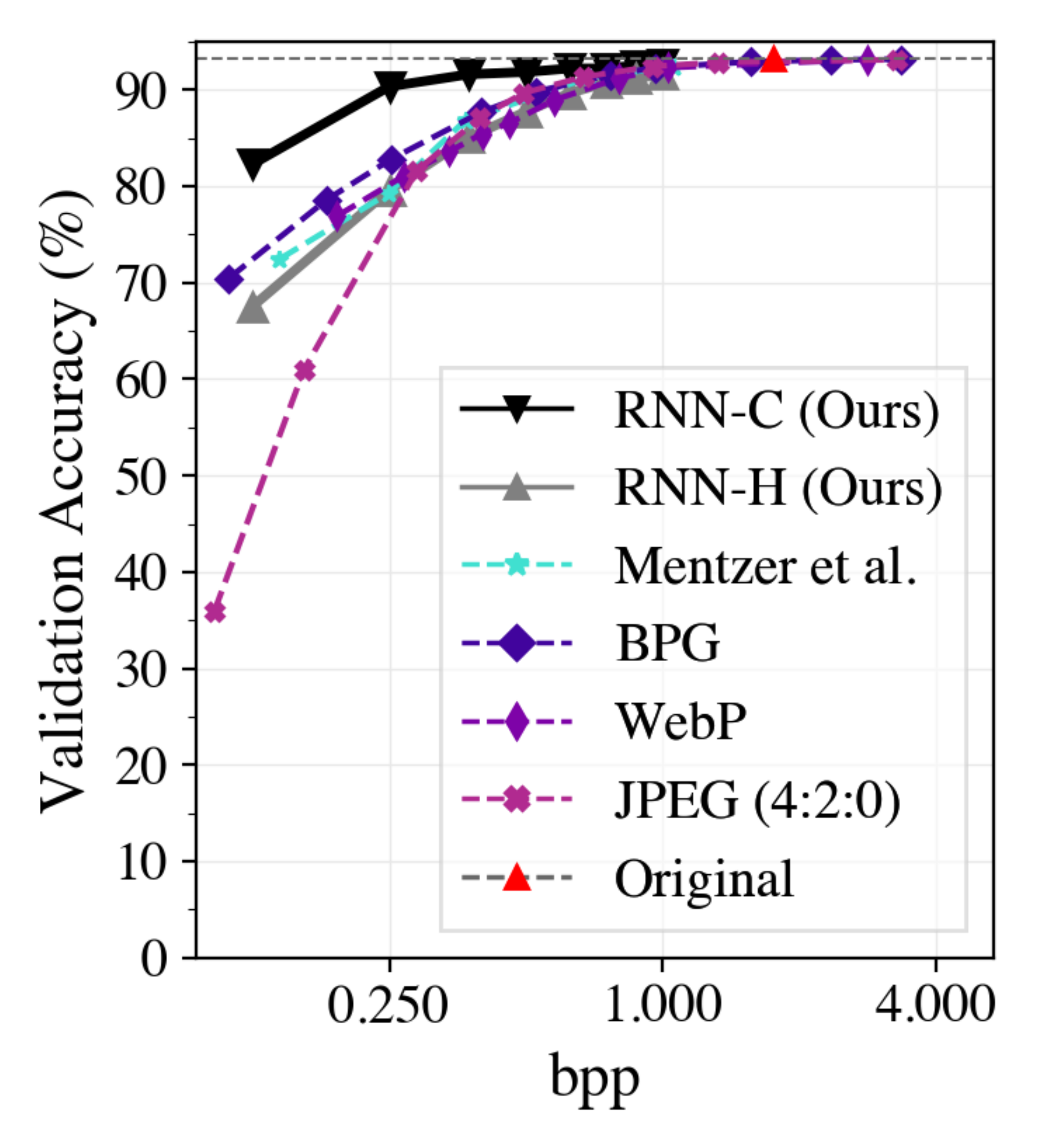}}%
        \hfill
        \subfigure[CUB-200-2011 Acc.]{\label{fig:accuracy_cub200}\includegraphics[width=.3\linewidth]{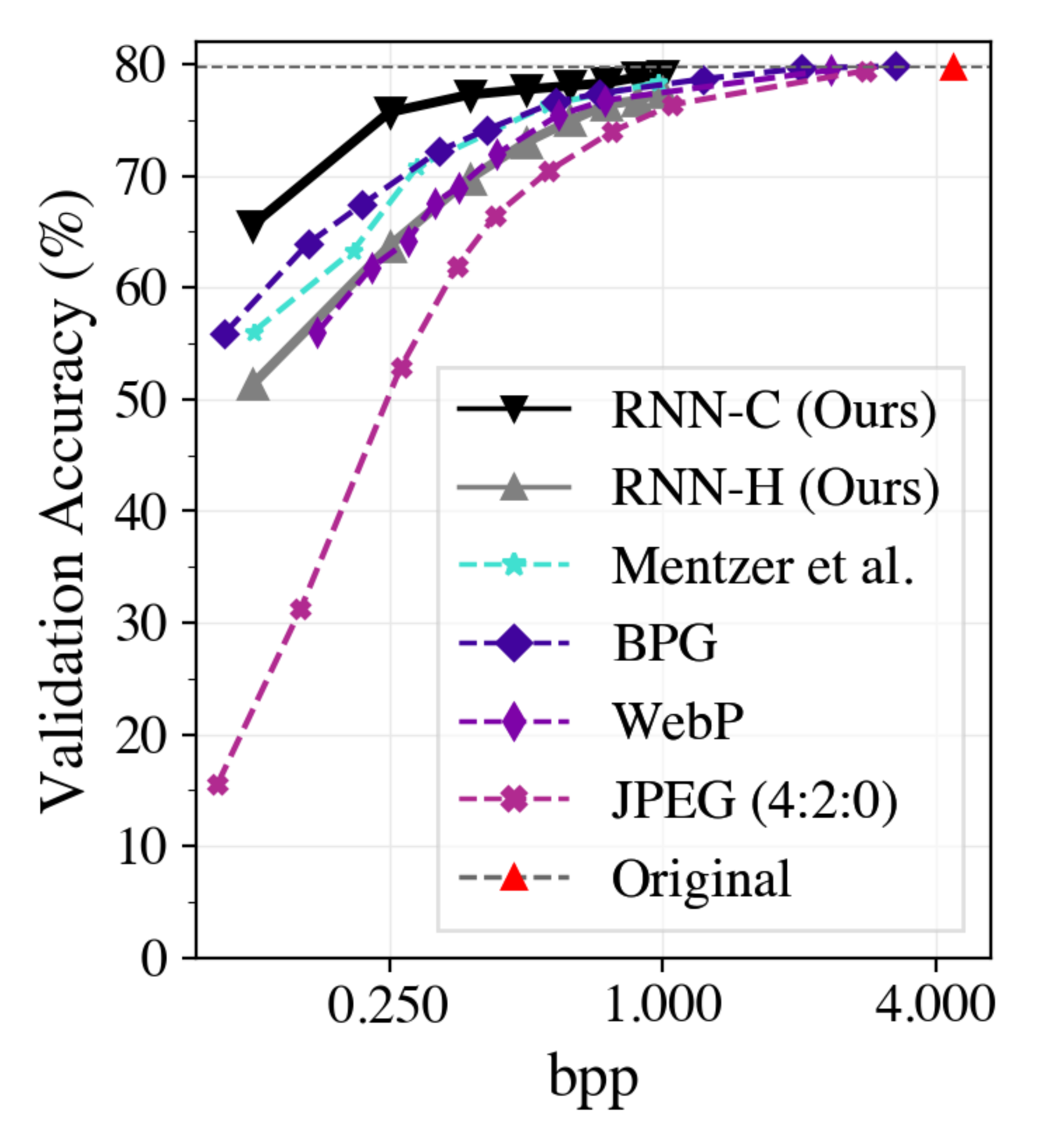}}
        \hfill
        \subfigure[MS-SSIM on Kodak]{\label{fig:ms_ssim_kodak}\includegraphics[width=.3\linewidth]{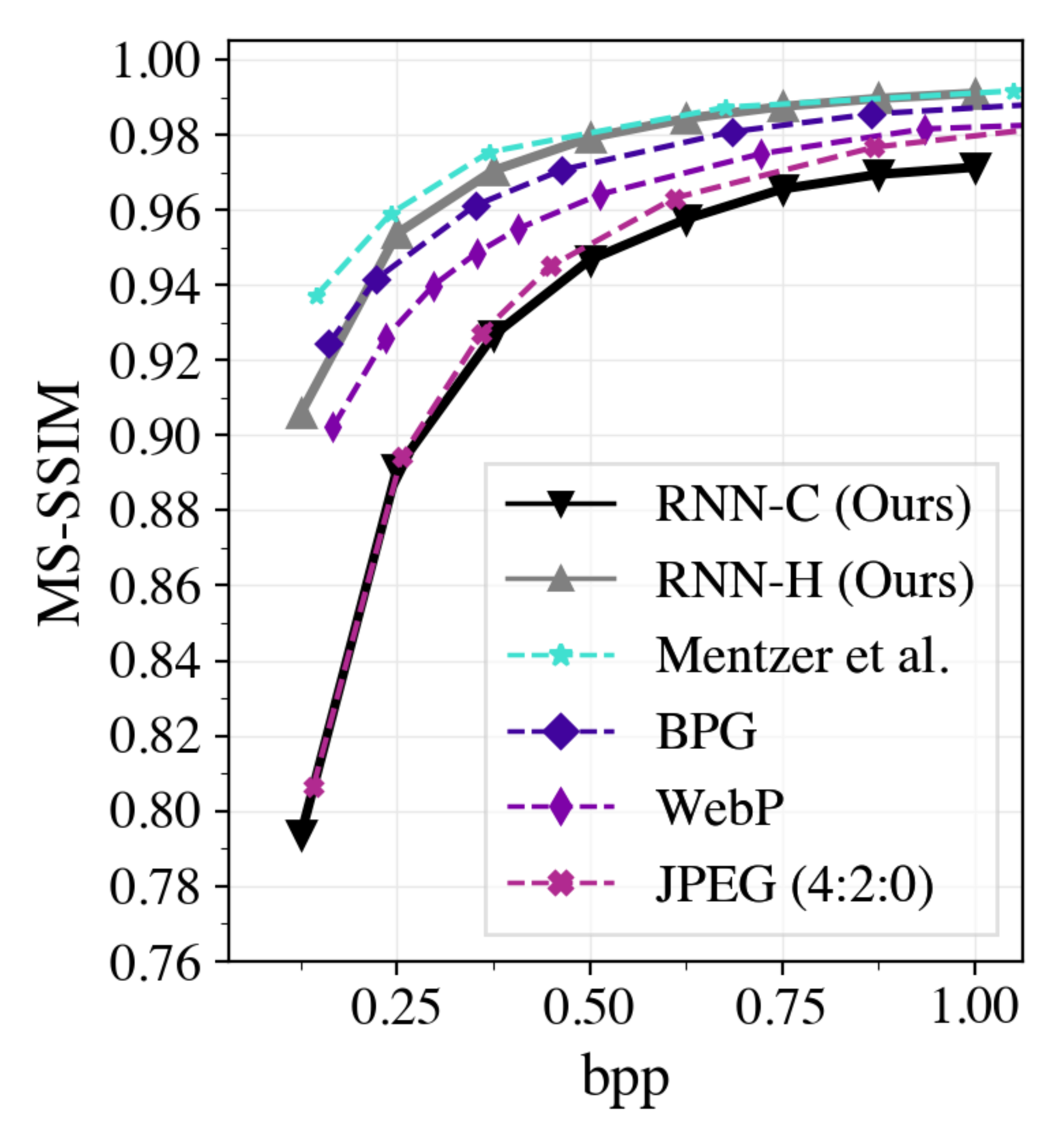}}%
        \caption{Validation accuracy is displayed in (a) and (b) where our RNN-C consistently outperforms BPG, JPEG, WebP, RNN-H and deep compression~\cite{mentzer2018} across bitrates. MS-SSIM on Kodak is shown in (c), indicating that our RNN-H is comptetitive to the state-of-the-art while RNN-C is comparable to JPEG.
        In each figure, RNN compression is trained from scratch on ImageNet-1K.}
    \end{figure}
    \paragraph{Fine-grained visual categorization.} In order to explore the generalization properties of our compression system to new tasks, we evaluate our method on two well known datasets for fine-grained visual categorization, namely Stanford Dogs and CUB-200-2011. We emphasize that the compression system is trained on the ImageNet-1K training split. Figures~\ref{fig:accuracy_dogs} and~\ref{fig:accuracy_cub200} indicate that RNN-C outperforms both the traditional codecs, RNN-H compression and the approach presented in \cite{mentzer2018} in terms of preserved classification accuracy with Inception-ResNet-V2 on both datasets. Similarly to ImageNet-1K classification, we see that the difference is especially pronounced below 0.5 bpp.
    \paragraph{Human visual perception.}
    \label{sec:results_human_perception}
    Figure~\ref{fig:ms_ssim_kodak} shows that RNN-H outperforms RNN-C on Kodak across bitrates. In comparison to neural compression from~\cite{mentzer2018}, RNN-H performs competitive, although worse for the lowest bitrate. Comparing our method against the traditional codecs, RNN-H slightly outperforms BPG for bitrates above 0.2 bpp. 
    Additionally, RNN-H clearly outperforms WebP and JPEG, while RNN-C performs competitive to JPEG.
    
    \section{Discussion}
    In this paper we investigate the trade-off in learned image compression with RNNs~\cite{toderici2017} between human visual perception and image classification. 
    To that end, we use a family of loss functions that enables us to either optimize compression for the human observer, or towards subsequent image classification.
    Our experiments show that when using the human friendly loss, RNN compression performs competitive to a state-of-the-art learned compression method~\cite{mentzer2018} and to the traditional codec BPG~\cite{bpg} in terms of MS-SSIM. 
    JPEG and WebP perform consistently worse than our approach. 
    We use MS-SSIM as a model for image similarity perceived by humans which, although being a widely adopted measure of distortion, is only an approximation to a true model of the human visual system. 
    Our classification friendly loss, based on features extracted from VGG-16, induces a compression system which by a large margin outperforms both the the other approaches in terms of preserved classification accuracy. 
    Our experiments furthermore indicate that our approach is agnostic to the CNN architecture used for classification and does not require the classifiers to be retrained on compressed images. 
    This suggests that we can indeed explicitly optimize image compression for subsequent classification.
    We observe a clear trade-off between quality perceived by the human visual system and classification accuracy, meaning that, for a fixed bitrate, an increase in accuracy always comes at the cost of degraded quality for the human observer, and vice versa. 
    Across classifiers, this trade-off is much more pronounced for bitrates below 0.5 bpp. 
    Finally, we find that by moving the loss function only marginally towards classification, we can substantially increase the preserved accuracy while incurring only a minor reduction in MS-SSIM. This improves compression in a scenario where images are consumed by humans and classifiers simultaneously and allows a user to trade off reconstruction quality accordingly.
    
    An interesting line of future work could include investigating other types of distortion measures used for the human oriented training loss, for example metrics that are based on CNNs which have been reported to correlate better with human perceptual similarity. Additionally, while classification is one of the most basic computer vision tasks, it would be interesting to explore whether the approach presented here also generalizes to other tasks such as image sementation and object detection.
    
    \section*{Acknowledgments}
    CZ and the DS3Lab gratefully acknowledge the support from the Swiss National Science Foundation (Project Number 200021\_184628), Innosuisse/SNF BRIDGE Discovery (Project Number 40B2-0\_187132), European Union Horizon 2020 Research and Innovation Programme (DAPHNE, 957407), Botnar Research Centre for Child Health, Swiss Data Science Center, Alibaba, Cisco, eBay, Google Focused Research Awards, Oracle Labs, Swisscom, Zurich Insurance, Chinese Scholarship Council, and the Department of Computer Science at ETH Zurich.

    \bibliographystyle{splncs04}
    \bibliography{refs}

\input{supp}

\end{document}

%% file: mathmacro.tex
\newcommand{\gI}{\mathcal{I}}

\newcommand{\gX}{\mathcal{X}}
\newcommand{\gY}{\mathcal{Y}}
\newcommand{\gZ}{\mathcal{Z}}

\newcommand\bR{\mathbb{R}}

\newcommand\bZ{\mathbb{Z}}

%% file: supp.tex
	\appendix
    
    \section{Choosing Layers for Reconstruction} As discussed in the main part, we conducted a series of preliminary experiments to find a choice of layers for the feature reconstruction loss. Since CNNs extract information at different levels of abstraction with different meaning in image space, the performance of a compression system trained to reconstruct features depends on the choice of layers. Recall that we use VGG-16 as loss network. Figure~\ref{fig:ablation_study} shows that when the goal is to train a compression system that generalizes across CNN architectures, the choice of layers that combines an early and a deep layer performs best across all bitrates. We suspect the reason for this observation is that when we only choose a deep layer or all layers, then the compression system overfits to the loss network and hence does not generalize to other architectures. If the goal is to mantain classification accuracy only for the loss network, Figure~\ref{fig:ablation_study_vgg16} shows that performance increases with depth and it is not necessary to include an early layer.
    \afterpage{%
    \begin{figure}[h!]
        \centering
        \subfigure[VGG-16]{\label{fig:ablation_study_vgg16}\includegraphics[width=0.32\linewidth]{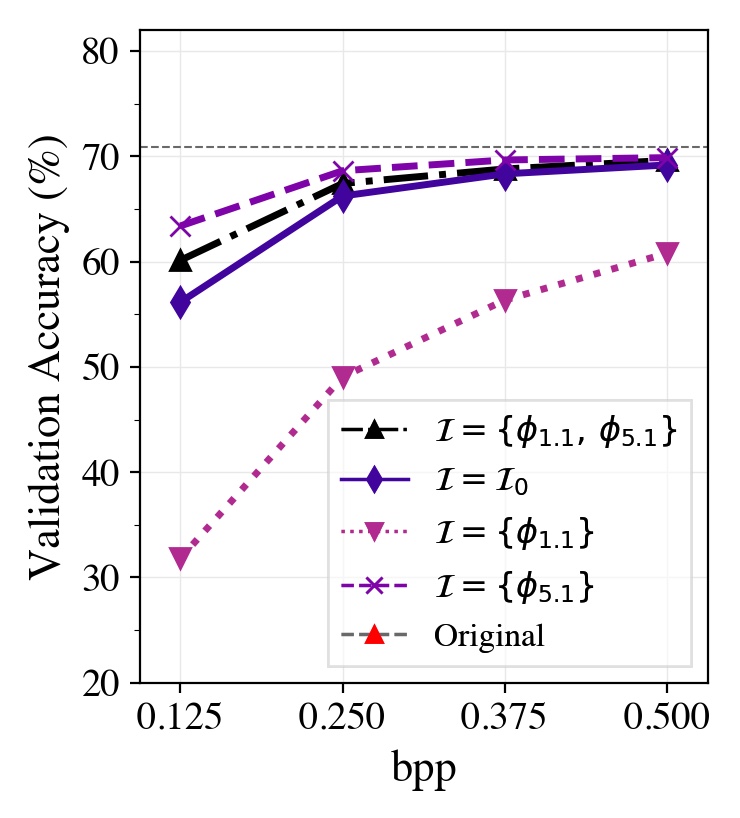}}%
        \hfill
        \subfigure[ResNet-50]{\includegraphics[width=0.32\linewidth]{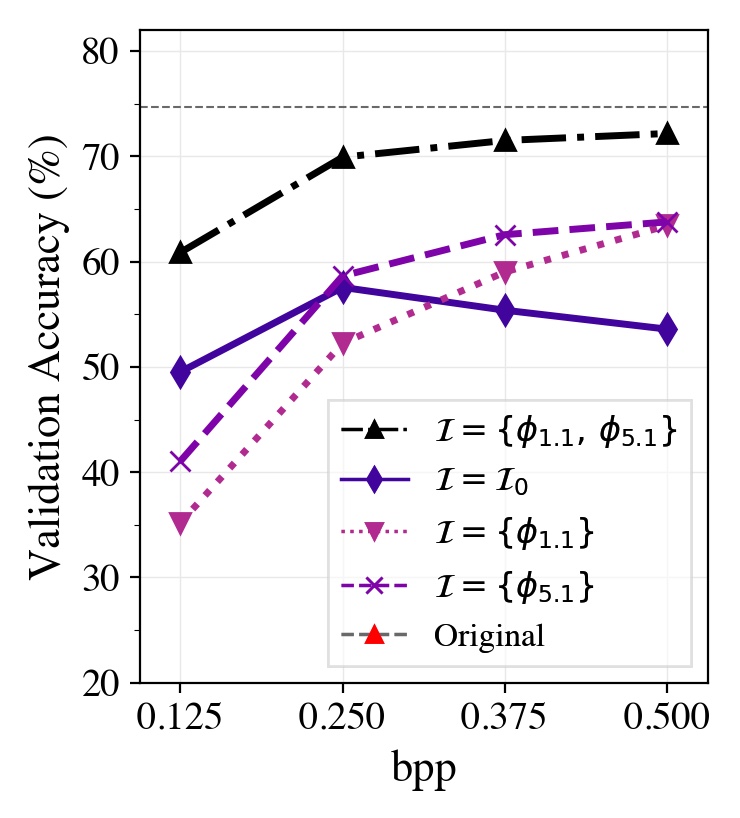}}%
        \hfill
        \subfigure[DenseNet-121]{\includegraphics[width=0.32\linewidth]{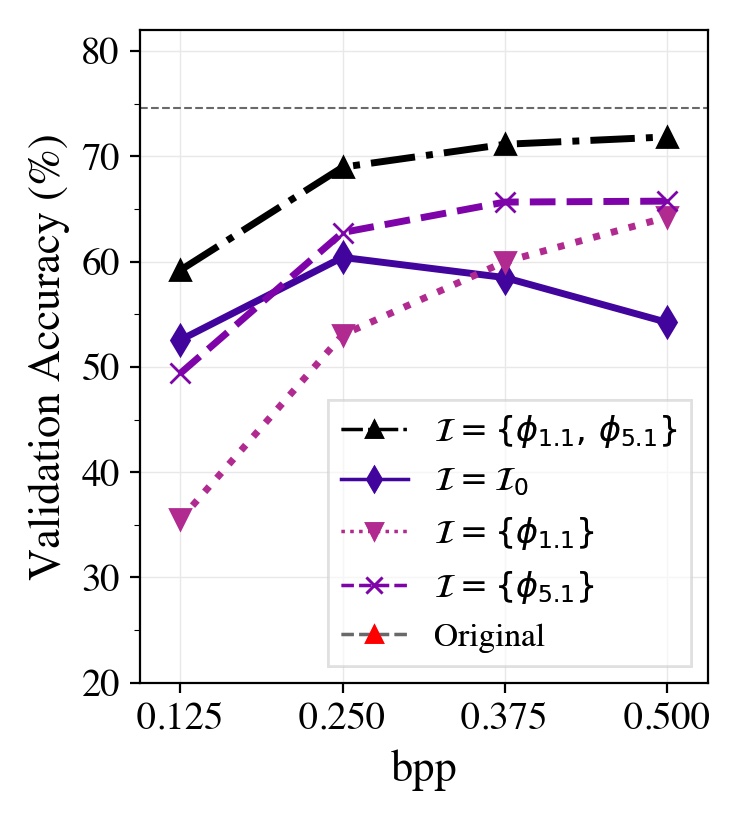}}
        \subfigure[MobileNet-V1]{\includegraphics[width=0.32\linewidth]{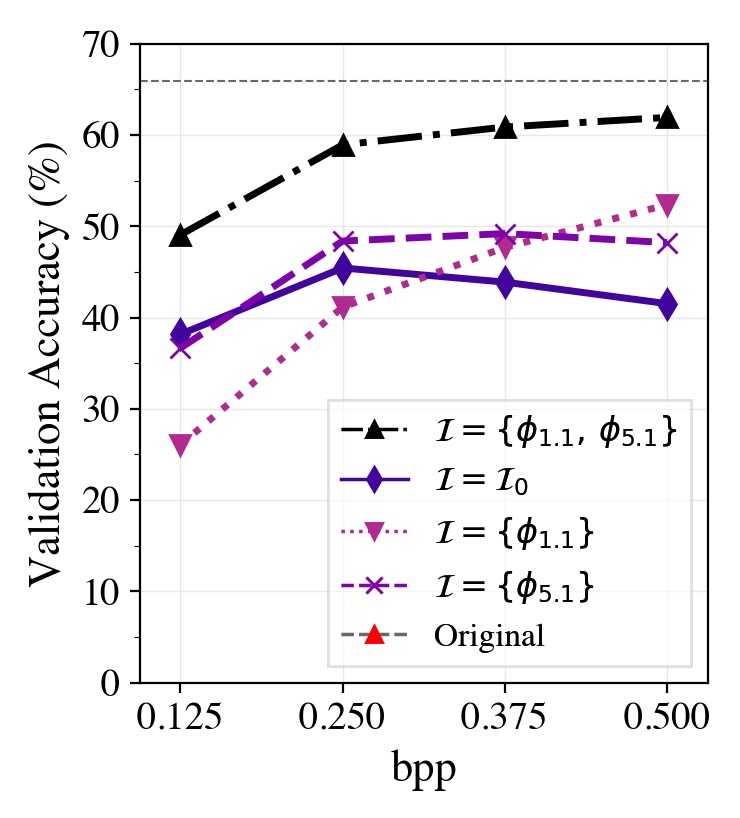}}%
        \hfill
        \subfigure[Inception-V3]{\includegraphics[width=0.32\linewidth]{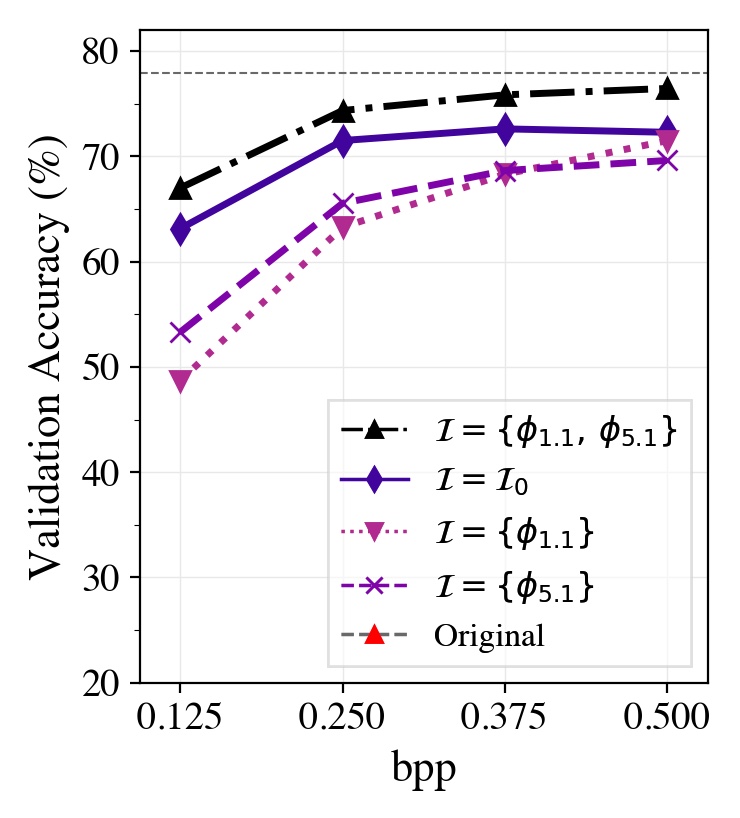}}%
        \hfill
        \subfigure[Inception-ResNet-V2]{\includegraphics[width=0.32\linewidth]{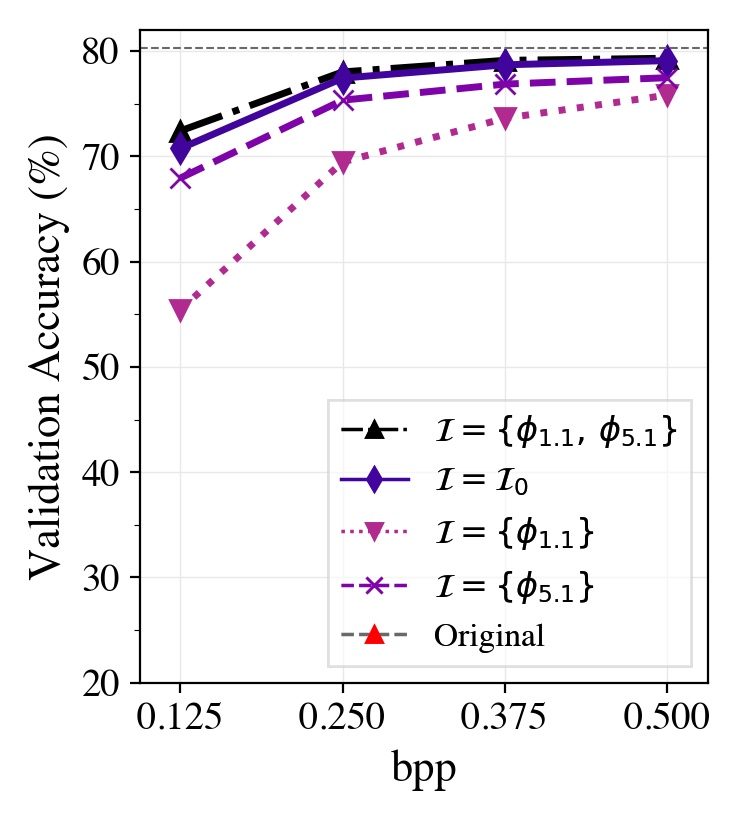}}
        \subfigure[Xception]{\includegraphics[width=0.32\linewidth]{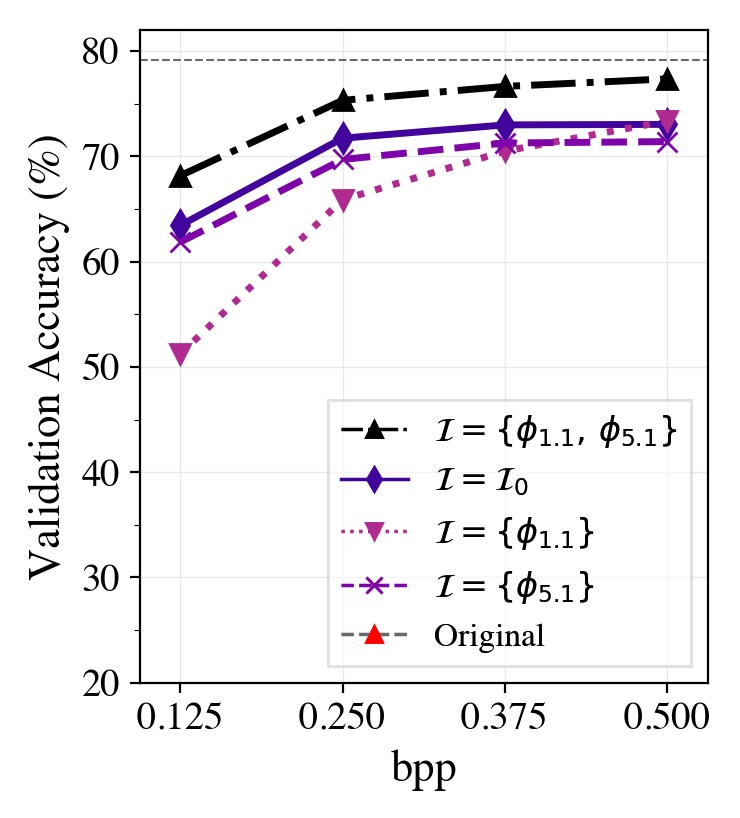}}
        \caption{Validation accuracy on ImageNet-1K for different layers in the feature reconstruction loss. \(\mathcal{I}_0\) corresponds to the full set of layers. VGG-16 is the loss network.}
        \label{fig:ablation_study}
    \end{figure}
    \clearpage
    }
    \newpage
    \section{MS-SSIM and Accuracy Plots} 
\begin{figure}
        \centering
        \includegraphics[width=.5\linewidth]{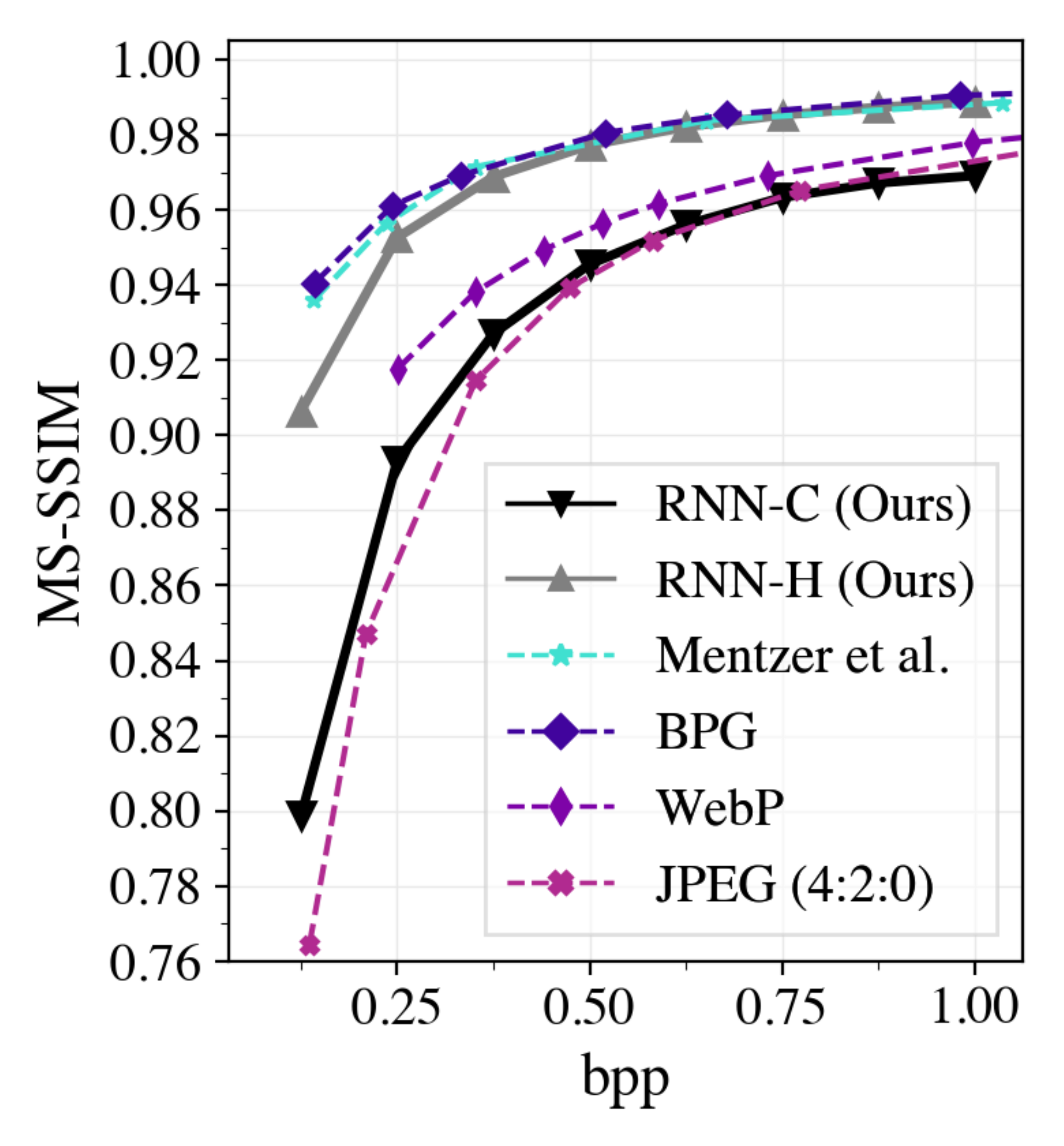}
        \caption{MS-SSIM on ImageNet-1K}
        \label{fig:ms_ssim_apx}
    \end{figure}

\begin{figure}
        \centering
        \subfigure[VGG-16]{\includegraphics[width=0.3\linewidth]{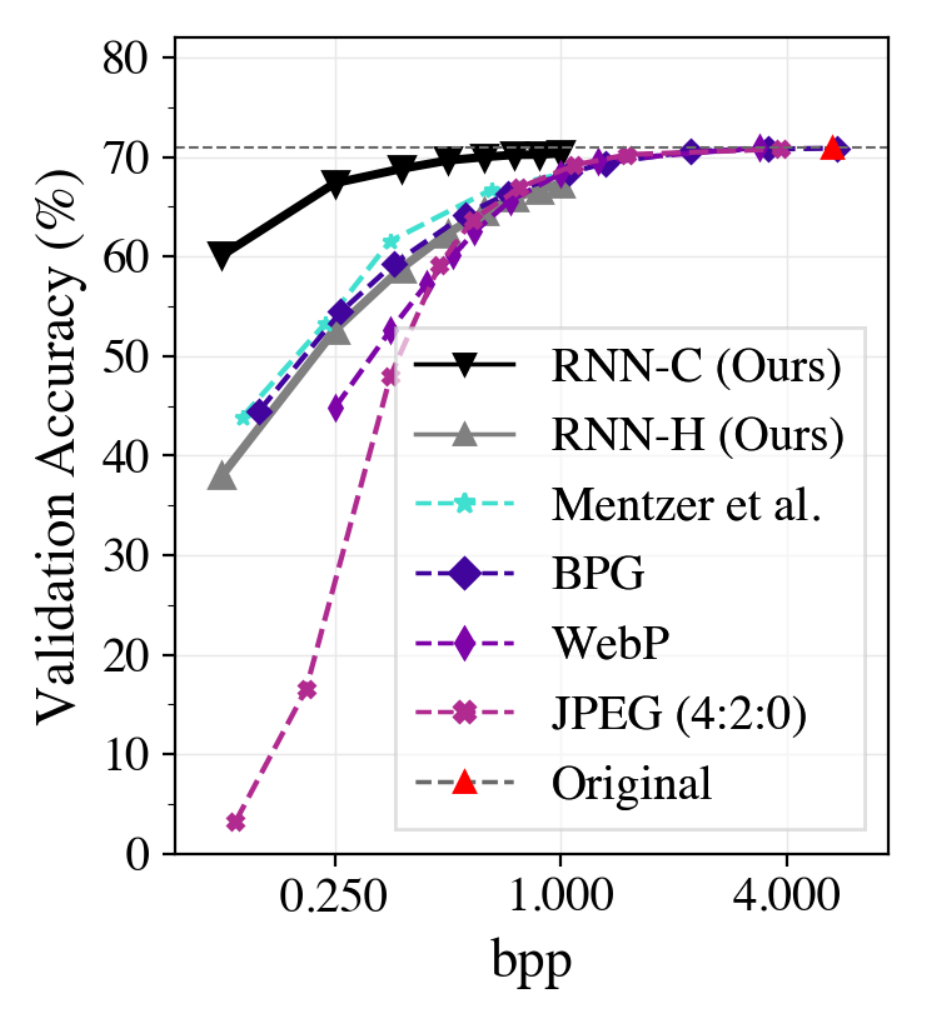}}%
        \hfill
        \subfigure[ResNet-50]{\includegraphics[width=0.3\linewidth]{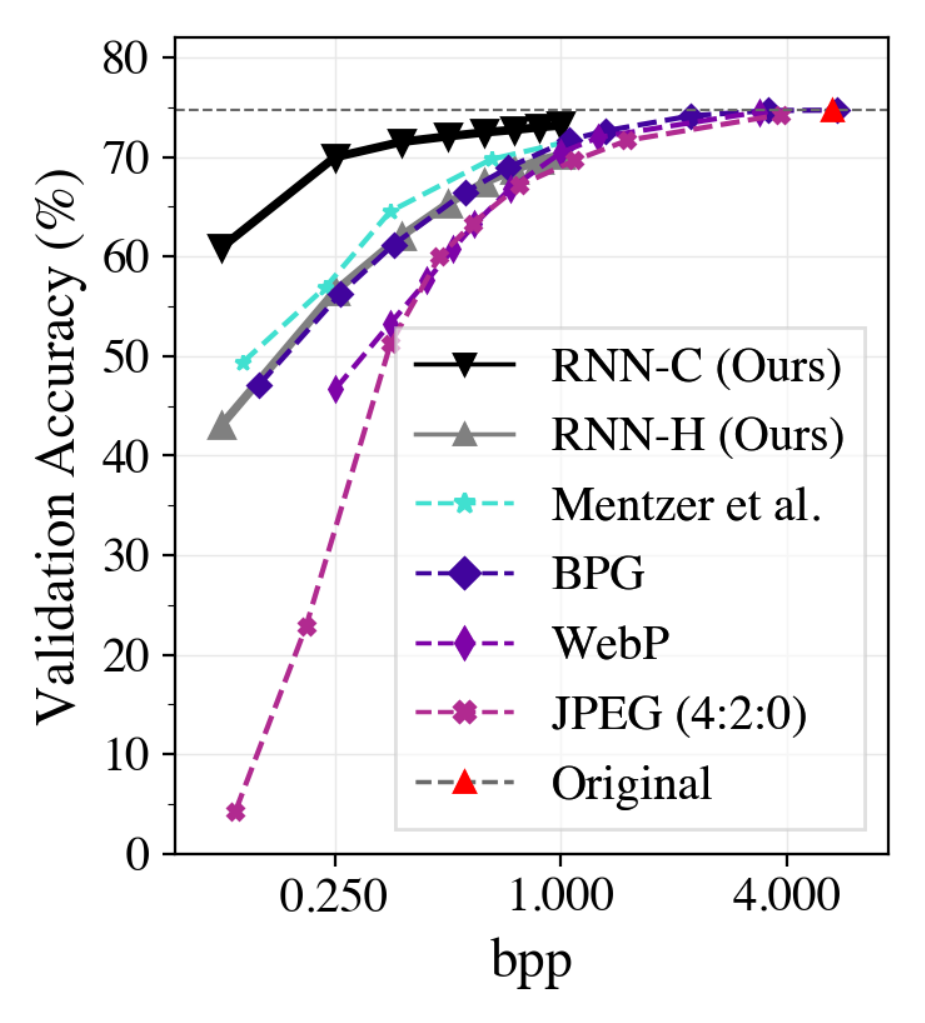}}%
        \hfill
        \subfigure[DenseNet-121]{\includegraphics[width=0.3\linewidth]{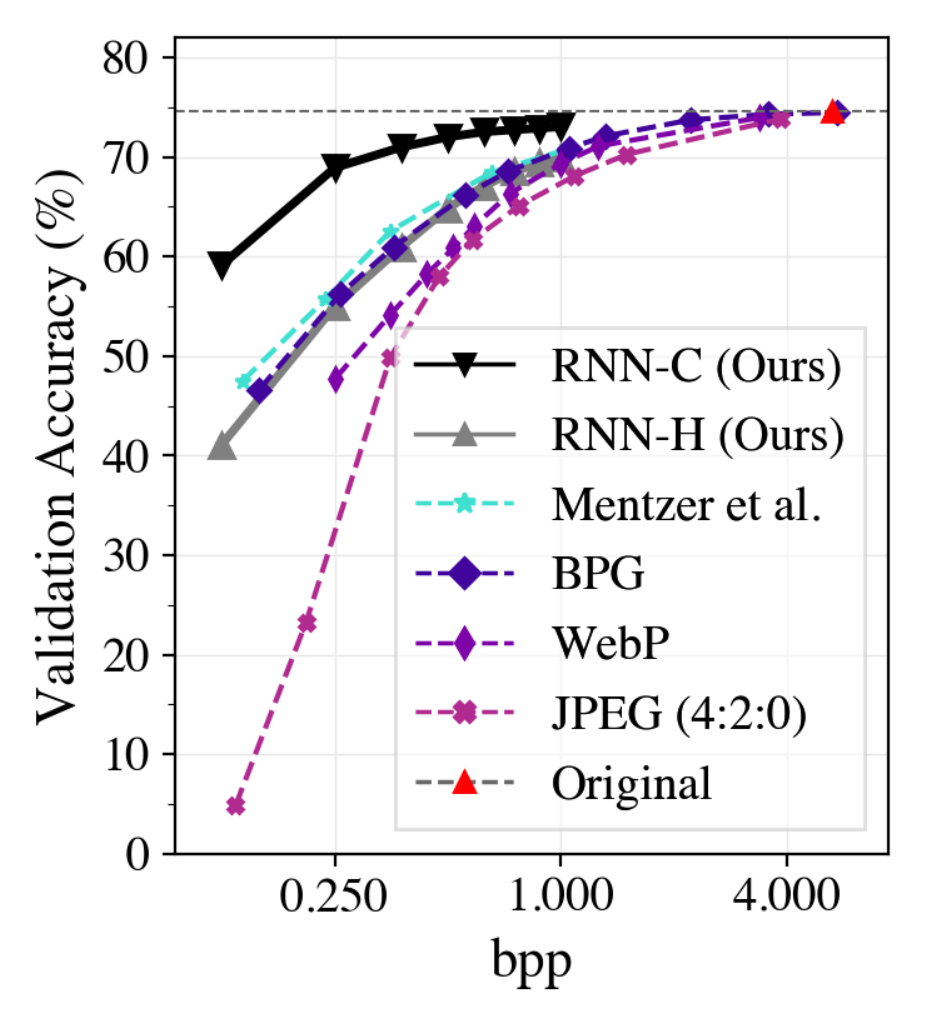}}
        \subfigure[MobileNet]{\includegraphics[width=0.3\linewidth]{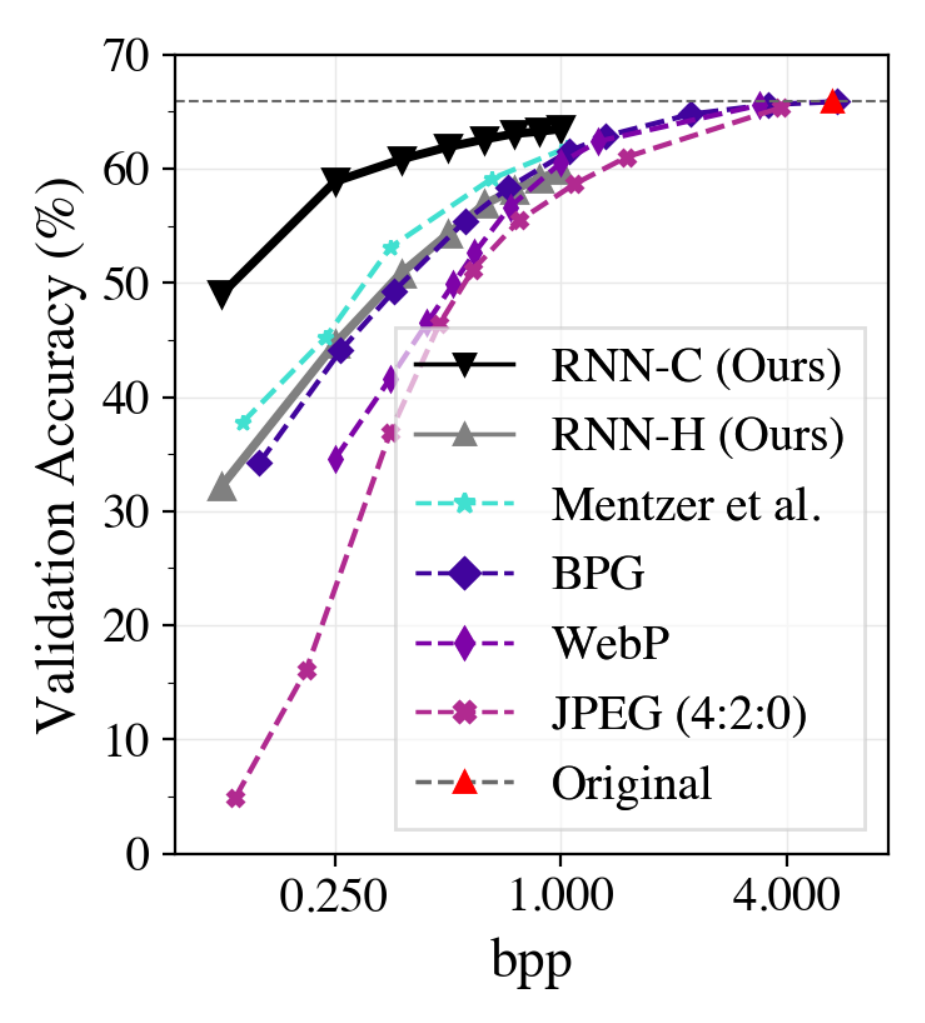}}%
        \hfill
        \subfigure[Inception-V3]{\includegraphics[width=0.3\linewidth]{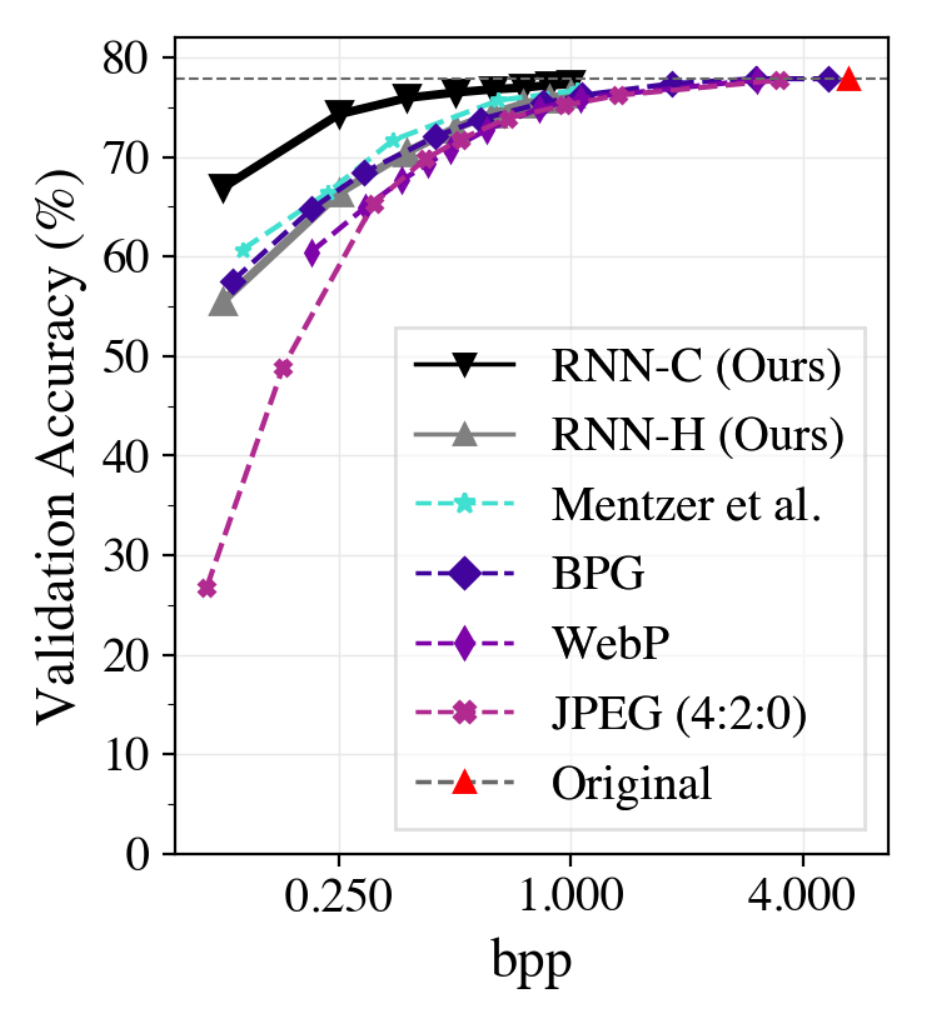}}%
        \hfill
        \subfigure[Xception]{\includegraphics[width=0.3\linewidth]{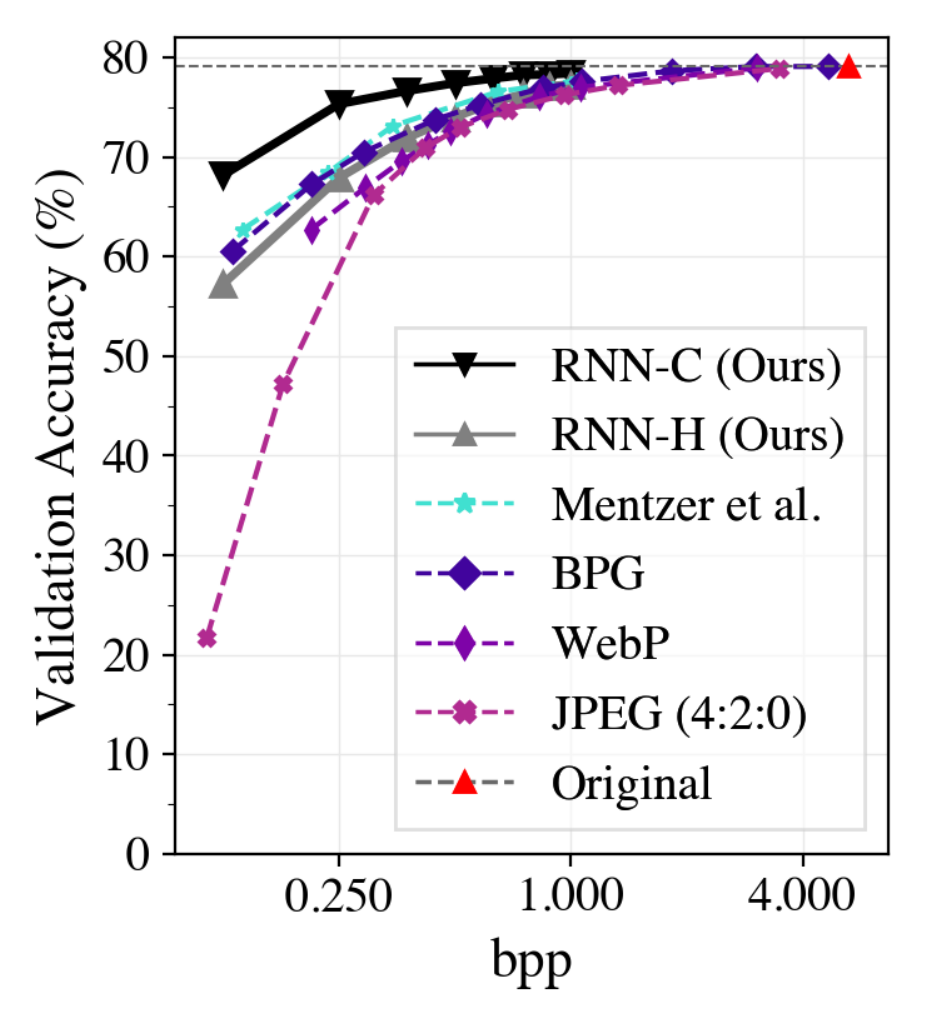}}%
        \hfill
        \subfigure[Inception-ResNet]{\includegraphics[width=.3\linewidth]{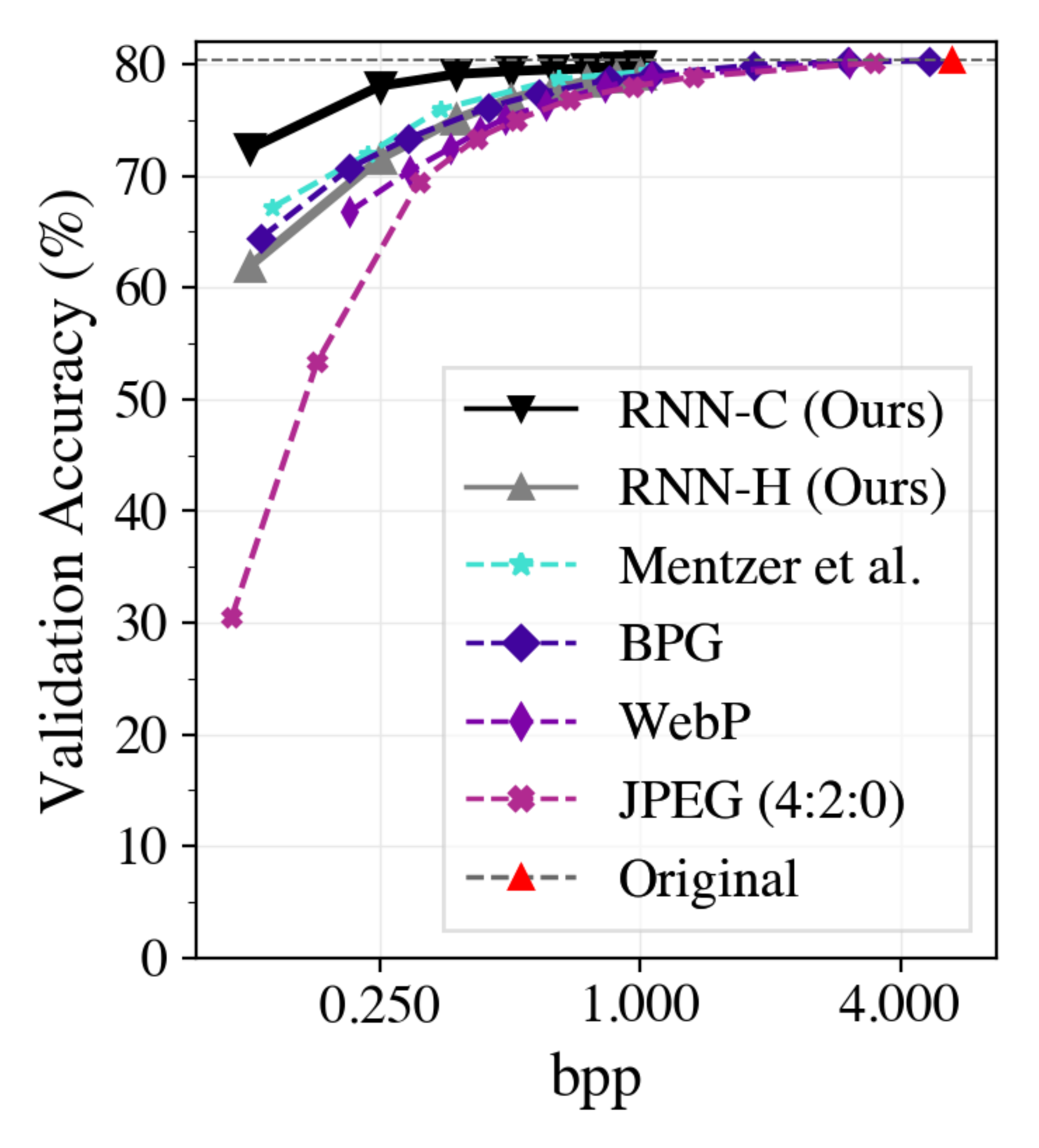}}
        \caption{Validation accuracy on ImageNet-1K for different classification architectures and compression algorithms.}
        \label{fig:accuracy_plots_imagenet}
    \end{figure}

\begin{figure}
        \centering
        \subfigure[VGG-16]{\includegraphics[width=0.45\linewidth]{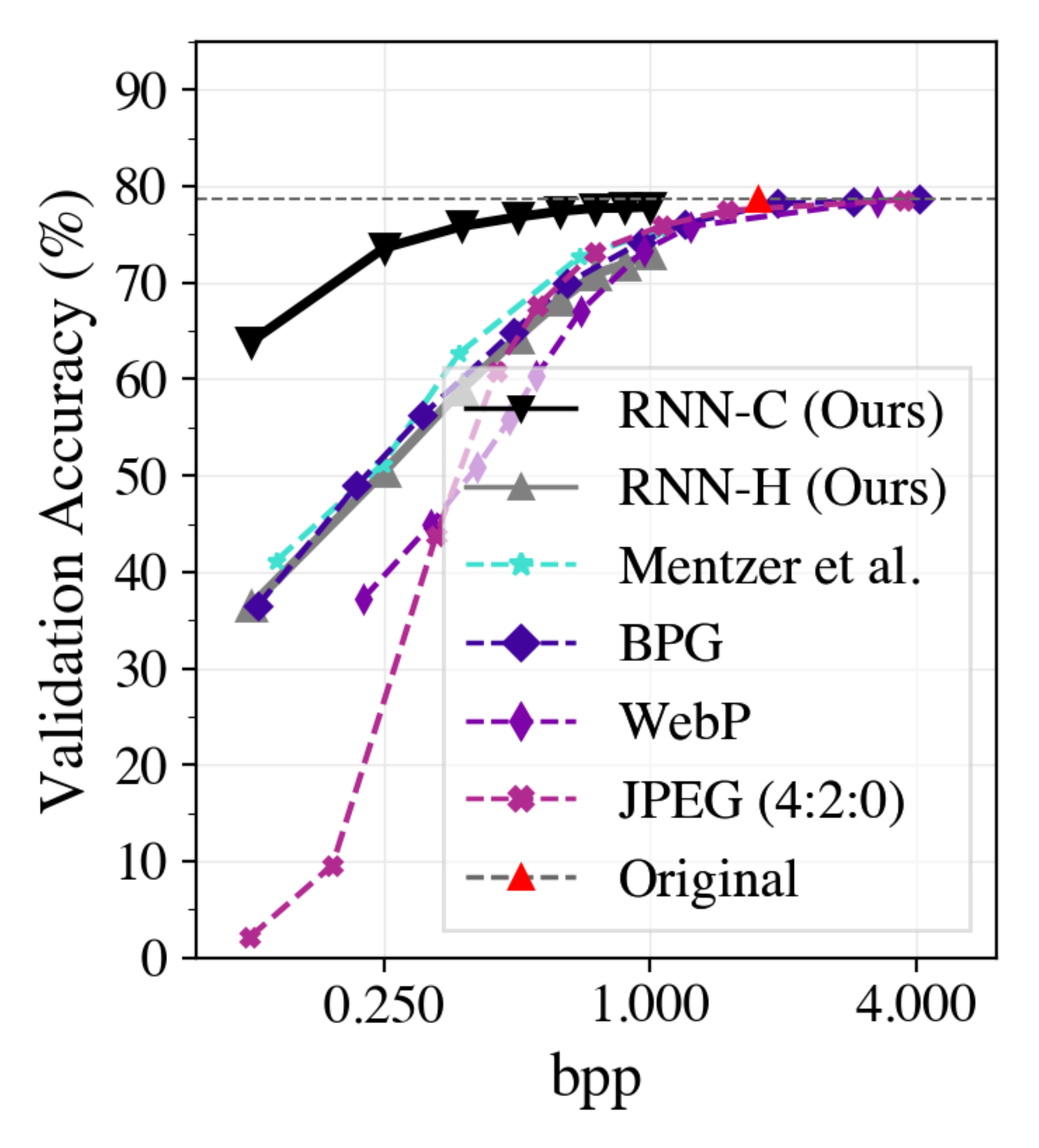}}%
        \hfill
        \subfigure[ResNet-50]{\includegraphics[width=0.45\linewidth]{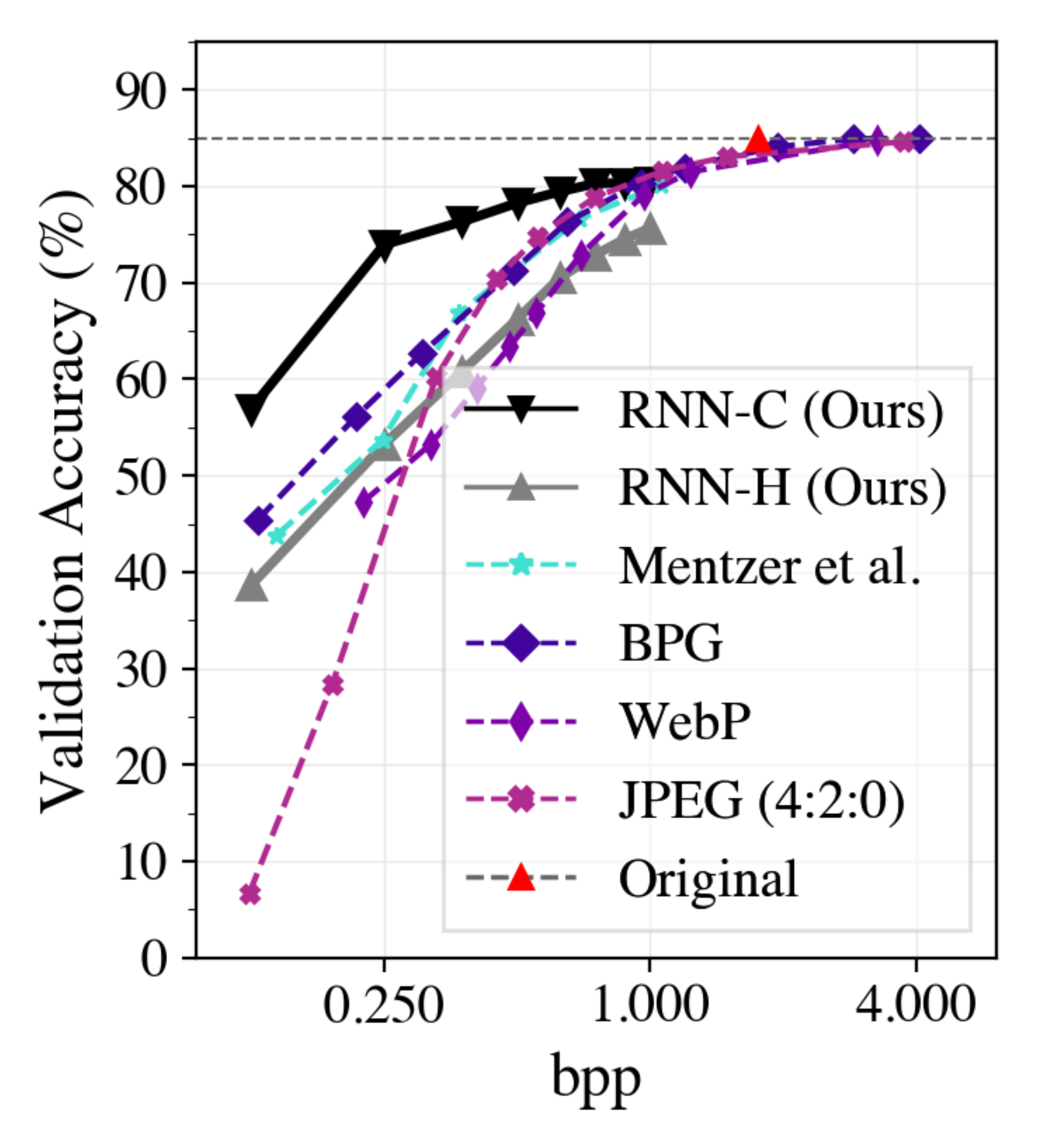}}
        \subfigure[MobileNet]{\includegraphics[width=0.45\linewidth]{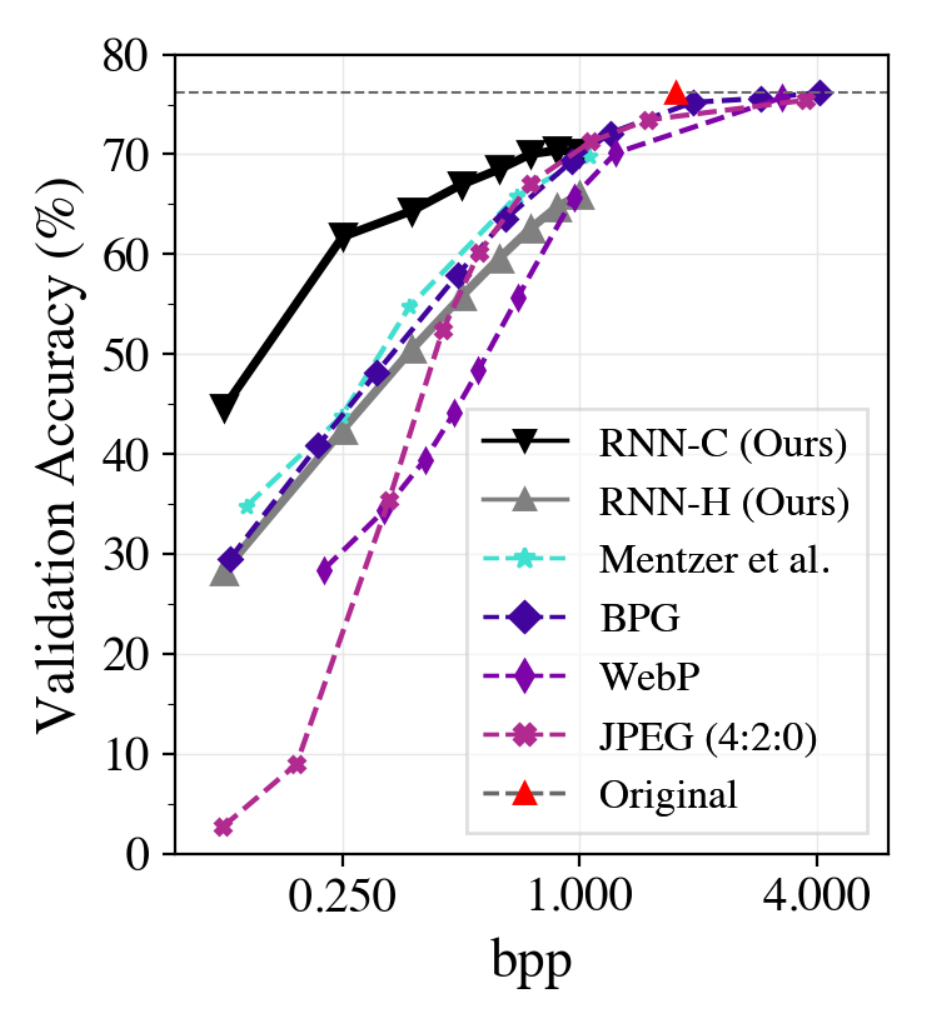}}%
        \hfill
        \subfigure[Inception-V3]{\includegraphics[width=0.45\linewidth]{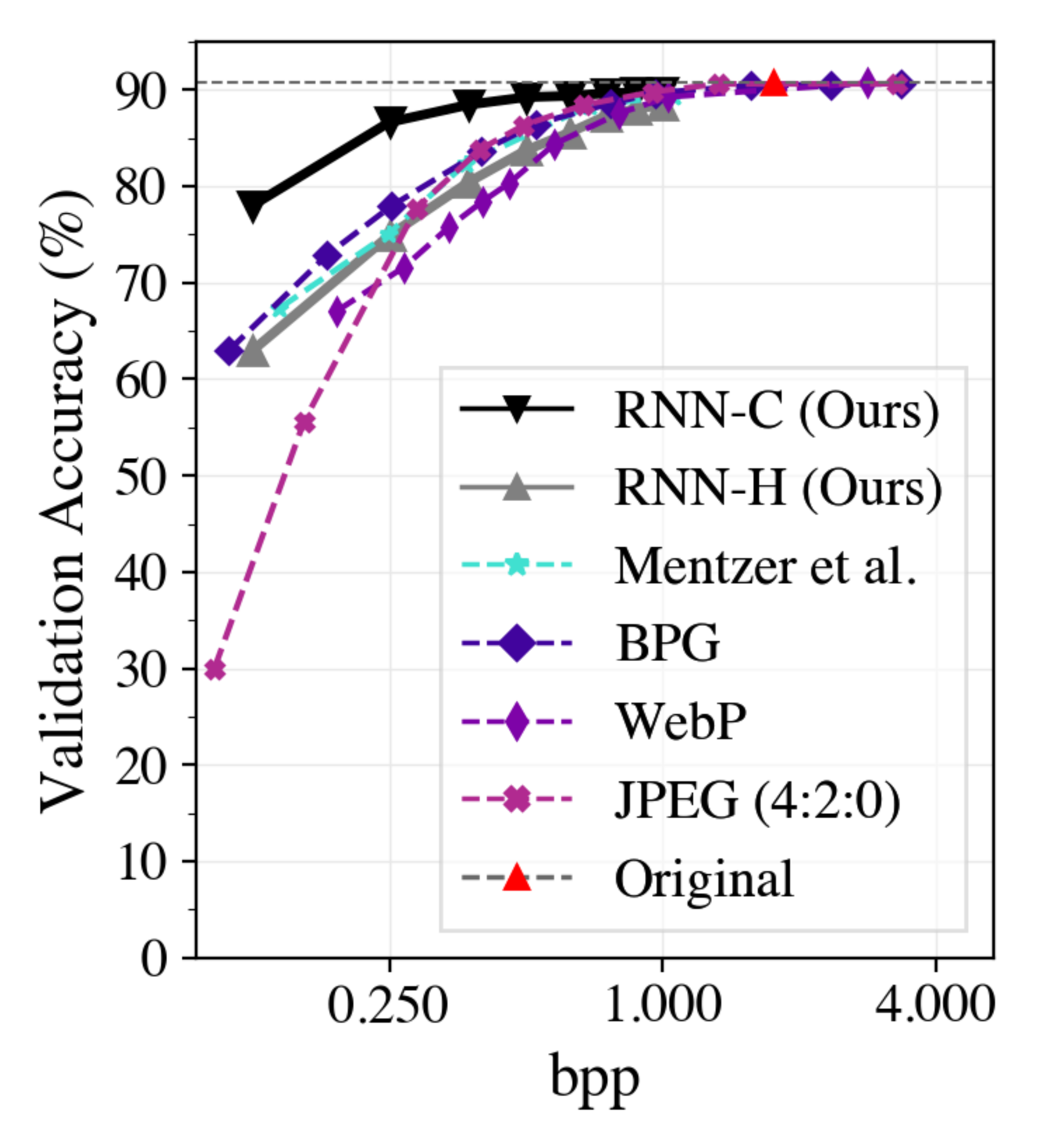}}
        \caption{Validation accuracy on Stanford Dogs for different classification architectures and compression algorithms.}
        \label{fig:accuracy_plots_stanford_dogs_apx}
    \end{figure}

\begin{figure}
        \centering
        \subfigure[VGG-16]{\includegraphics[width=0.45\linewidth]{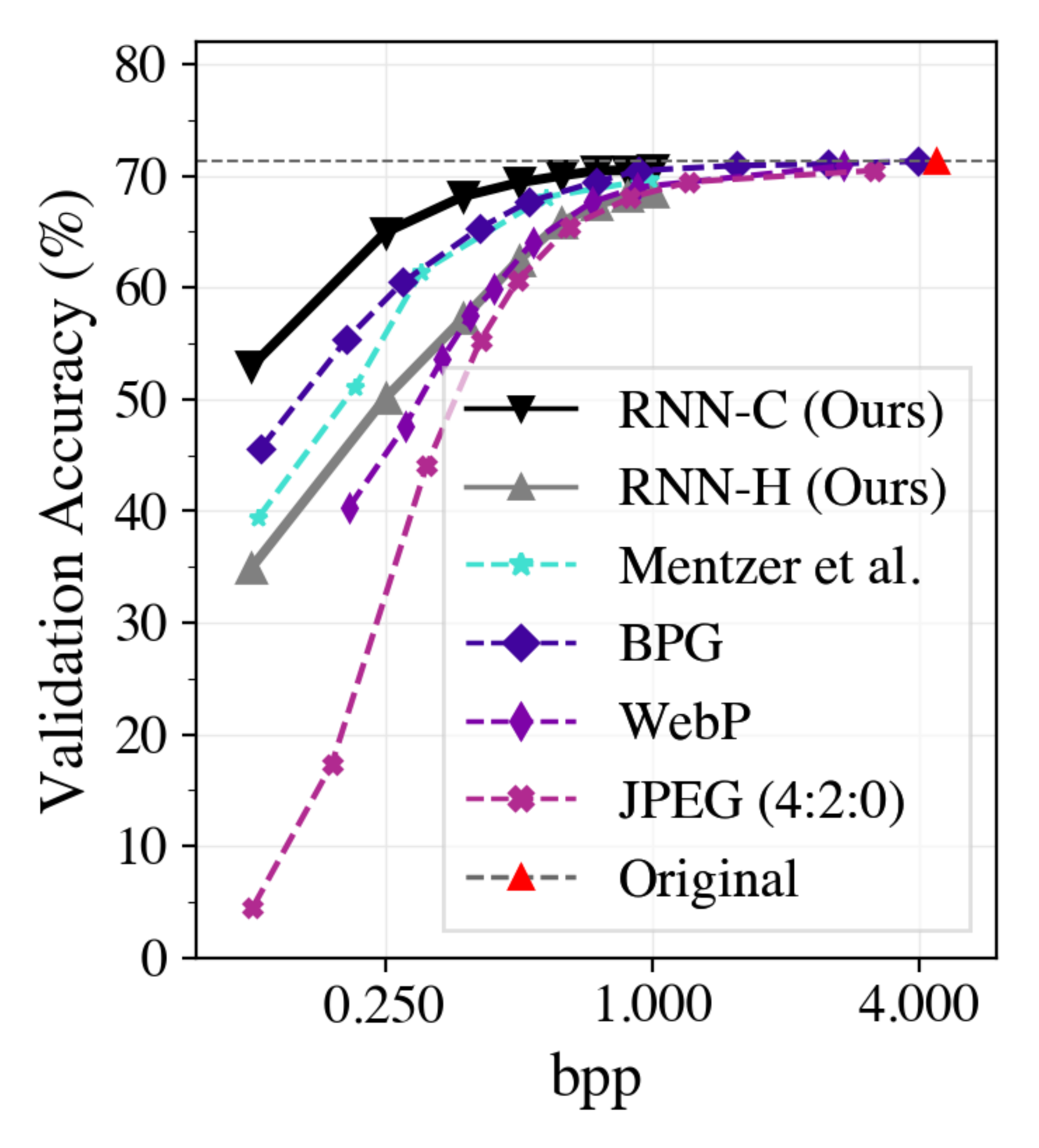}}%
        \hfill
        \subfigure[ResNet-50]{\includegraphics[width=0.45\linewidth]{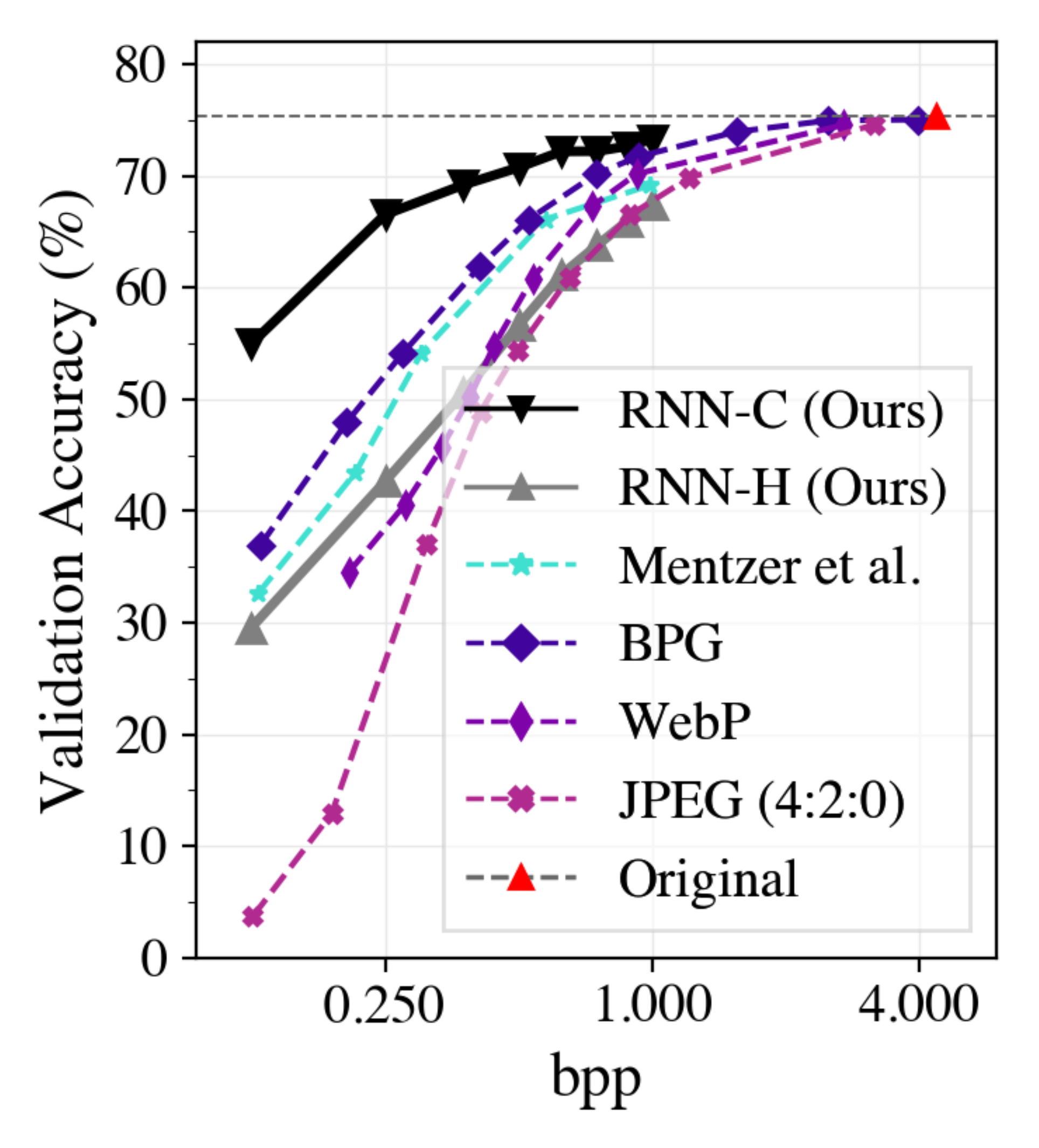}}
        \subfigure[MobileNet]{\includegraphics[width=0.45\linewidth]{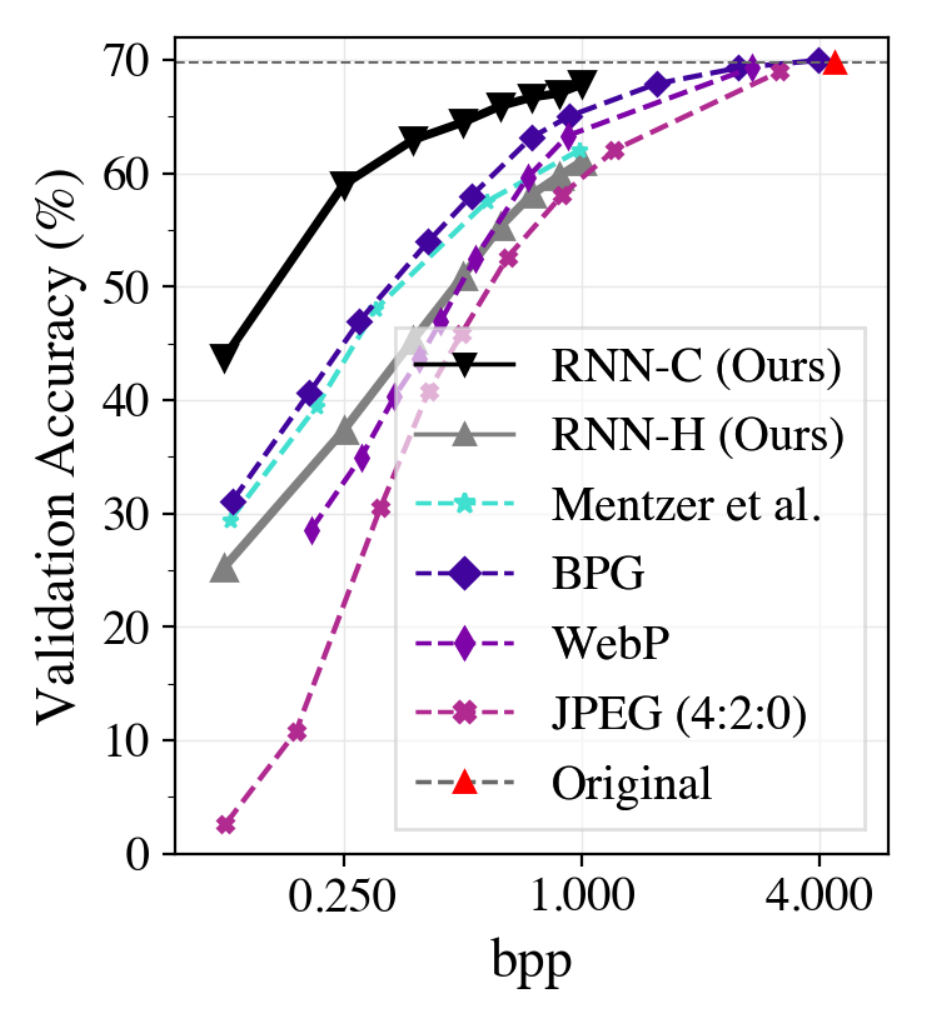}}%
        \hfill
        \subfigure[Inception-V3]{\includegraphics[width=0.45\linewidth]{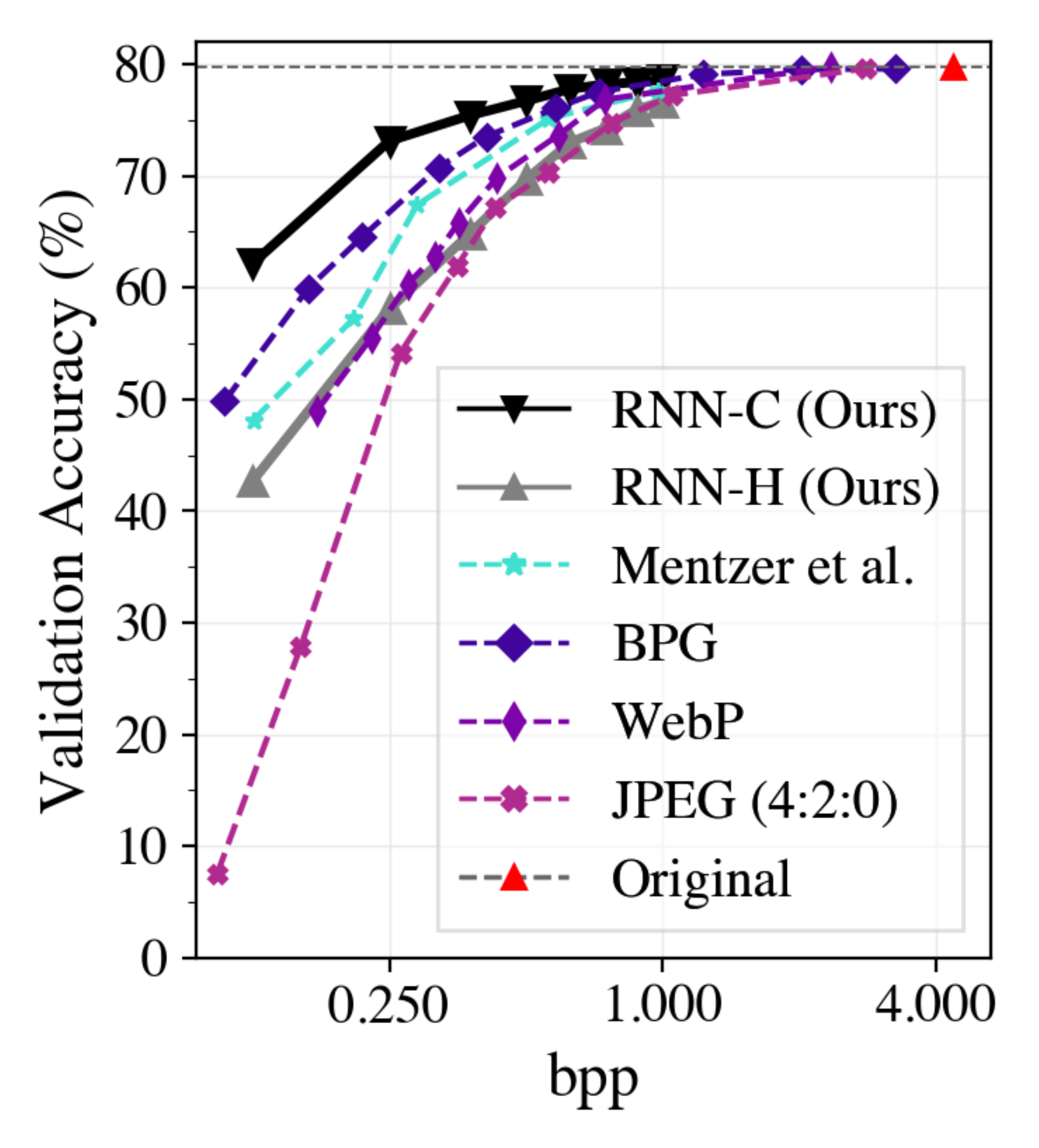}}
        \caption{Validation accuracy on CUB-200-2001 for different classification architectures and compression algorithms.}
        \label{fig:accuracy_plots_cub200_apx}
    \end{figure}

    Figure~\ref{fig:ms_ssim_apx} show the MS-SSIM scores on the validation set of ImageNet-1K. It can be seen that BPG outperforms RNN-H with respect to MS-SSIM, especially for the lowest bitrates. RNN-H clearly outperforms WebP and JPEG, while RNN-C performs competitive to JPEG.
    Figure~\ref{fig:accuracy_plots_imagenet} visualizes our findings from the main part on the ImageNet-1K validation set. We see that, across classifiers and bitrates, RNN-C performs best in terms of preserved classification accuracy. Figures~\ref{fig:accuracy_plots_stanford_dogs_apx} and~\ref{fig:accuracy_plots_cub200_apx} indicate the same trend for the fine grained visual categorization experiments.
    
    
    \newpage
    \section{Success and Failure Cases}
    \begin{figure}
        \centering
        \subfigure[BPG]{\includegraphics[width=.22\linewidth]{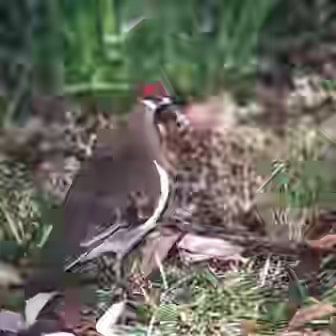}}%
        \hfill
        \subfigure[RNN-C]{\includegraphics[width=.22\linewidth]{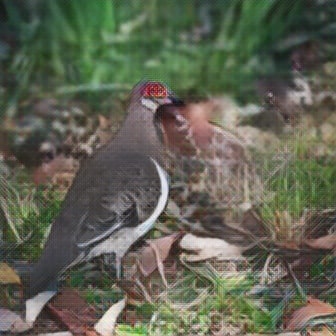}}%
        \hfill
        \subfigure[BPG]{\includegraphics[width=.22\linewidth]{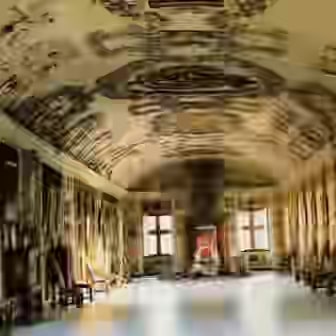}}%
        \hfill
        \subfigure[RNN-H]{\includegraphics[width=.22\linewidth]{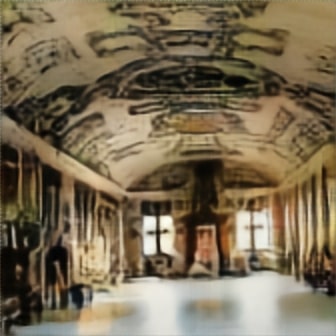}}
        \caption{Images (compressed to $\sim0.1$bpp) where our methods are better than BPG in terms of classification (left) and MS-SSIM (right). On the left, Inception-ResNet predicts ``Partridge" on the original and RNN-C image and ``Black Grouse" on the BPG compressed image. On the right, BPG achieves 0.88 MS-SSIM and RNN-H 0.89.}
        \label{fig:success_cases}
    \end{figure}
    \begin{figure}
        \centering
        \subfigure[BPG]{\label{fig:failure_classification1}\includegraphics[width=.24\linewidth]{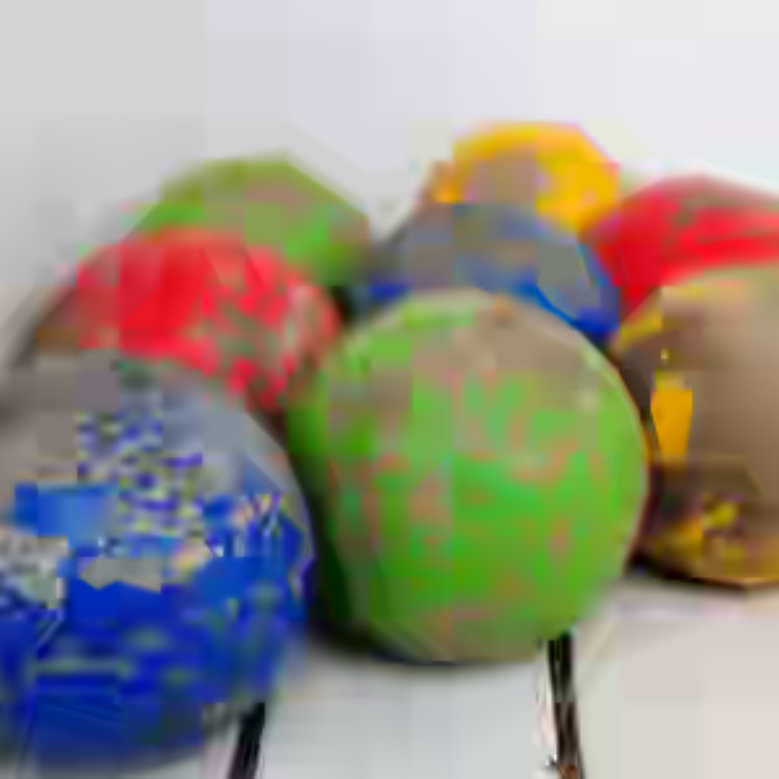}}%
        \hfill
        \subfigure[RNN-C]{\label{fig:failure_classification2}\includegraphics[width=.24\linewidth]{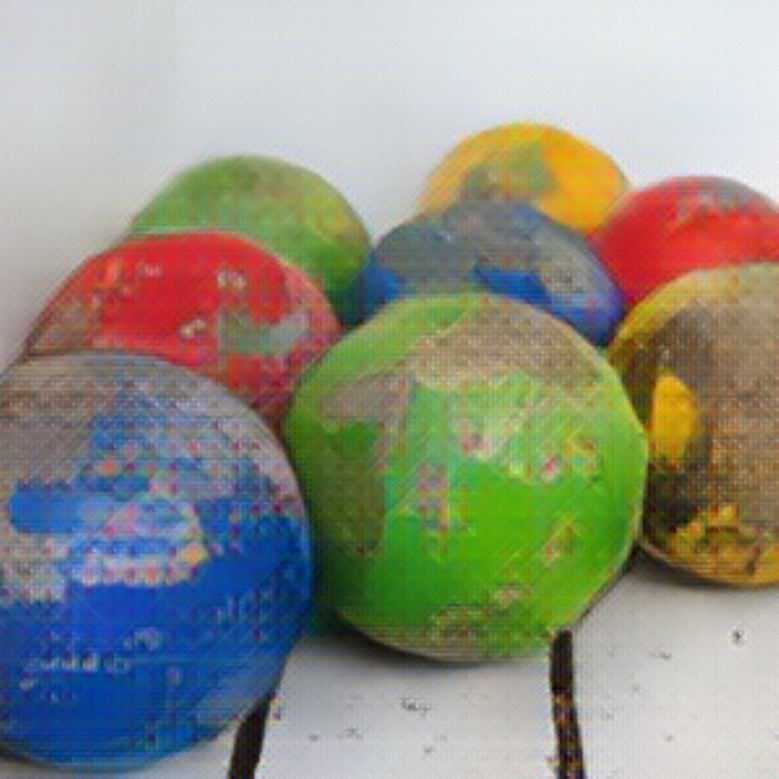}}%
        \hfill
        \subfigure[BPG]{\label{fig:failure_human1}\includegraphics[width=.18\linewidth]{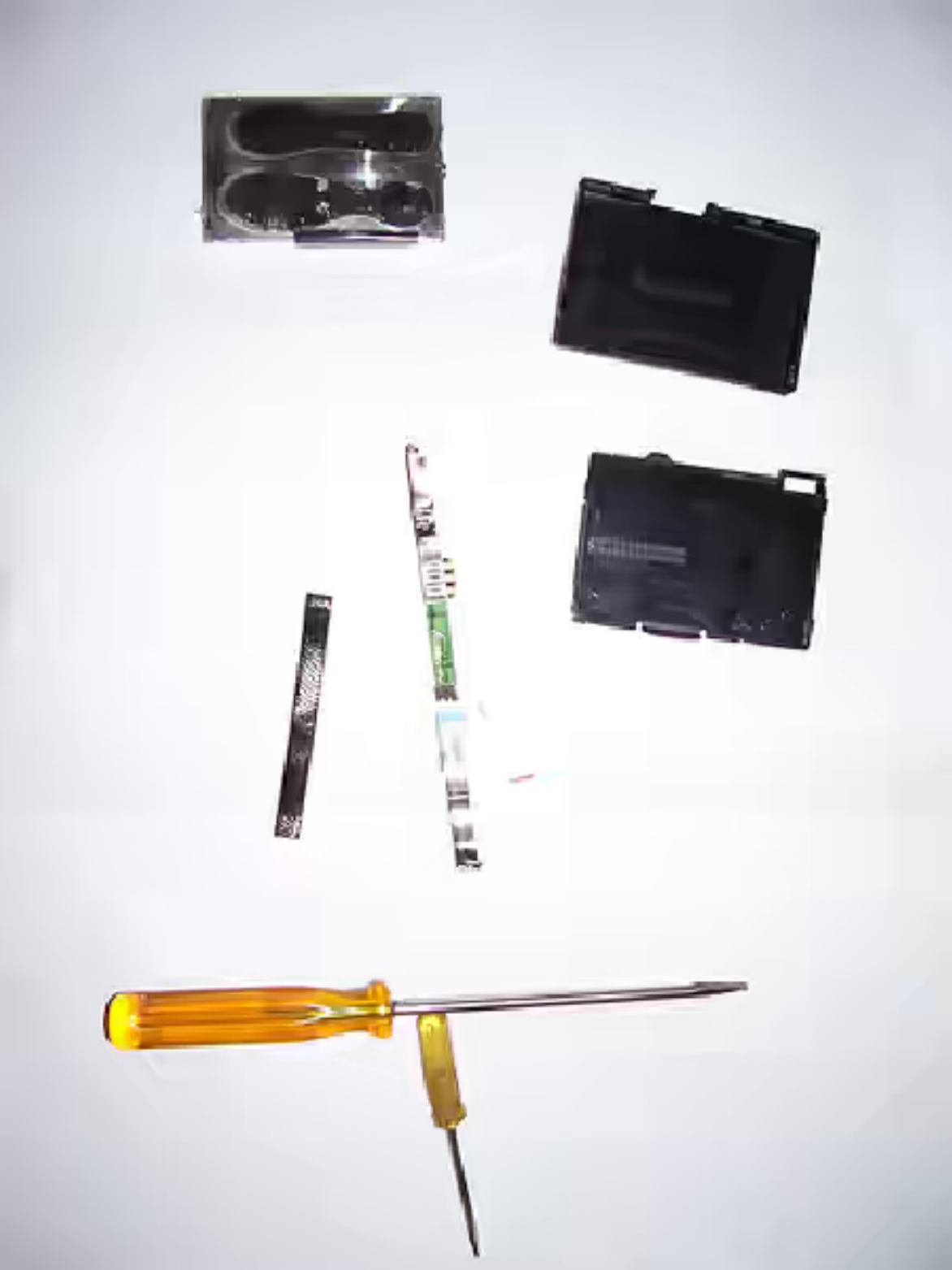}}%
        \hfill
        \subfigure[RNN-H]{\label{fig:failure_human2}\includegraphics[width=.18\linewidth]{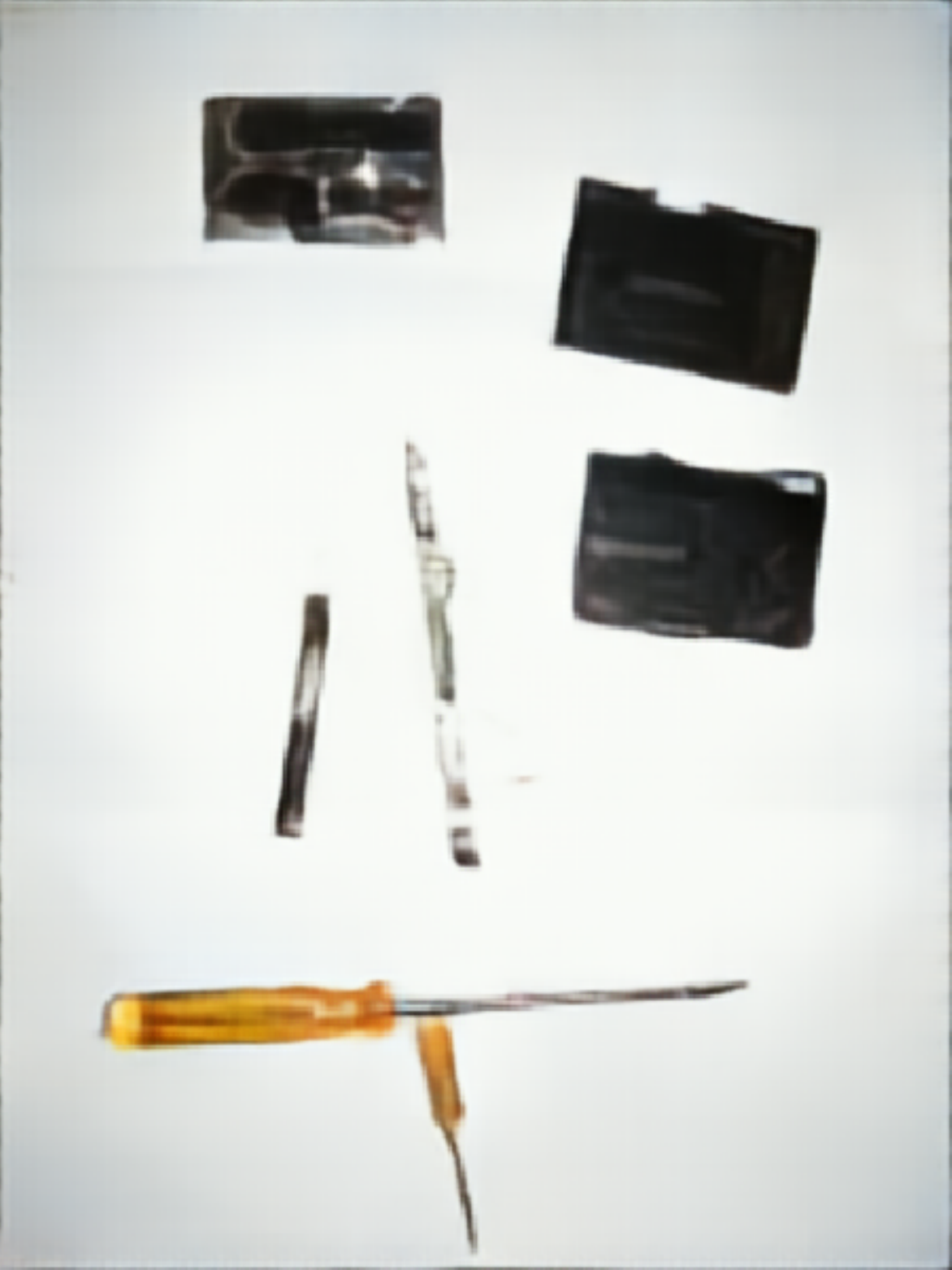}}%
        \caption{Images (compressed to $\sim0.1$bpp) where our methods are worse than BPG in terms of classification (left) and MS-SSIM (right). Inception-ResNet predicts ``Croquet Ball" on the original and BPG image and ``Abacus" on the RNN-C compressed image. In the right, BPG achieves 0.99 MS-SSIM and RNN-H 0.98.}
        \label{fig:failure_cases}
    \end{figure}
    
    \newpage
    \section{Visual Examples} 
\begin{figure}
        \centering
        \subfigure[\textbf{Original}]{\includegraphics[width=.49\linewidth]{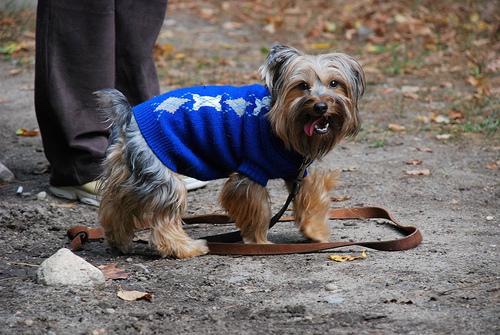}}%
        \hfill
        \subfigure[\textbf{RNN-H}, 0.125 bpp]{\label{fig:terrier_rnn_h}\includegraphics[width=.49\linewidth]{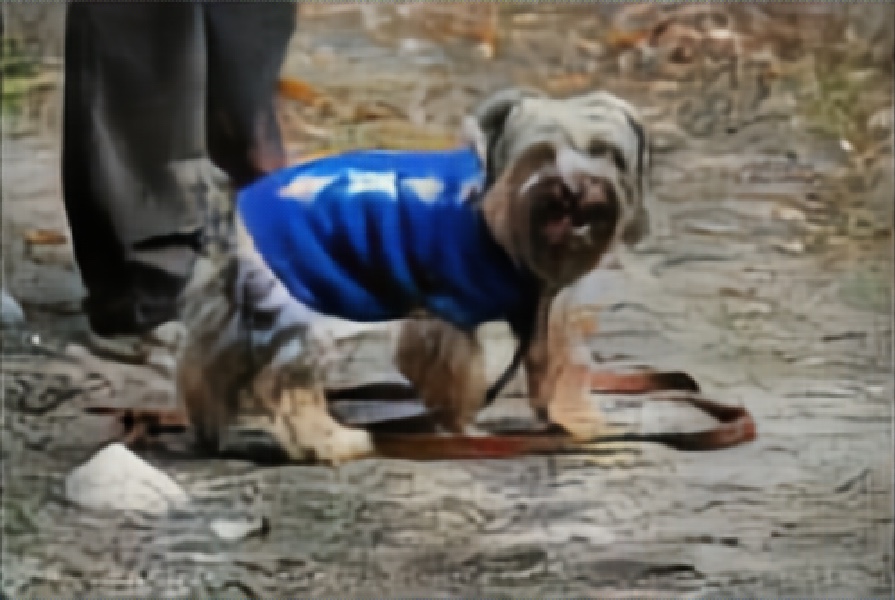}}
        \subfigure[\textbf{Mentzer \etal}~\cite{mentzer2018}, 0.177 bpp]{\label{fig:terrier_cpm}\includegraphics[width=.49\linewidth]{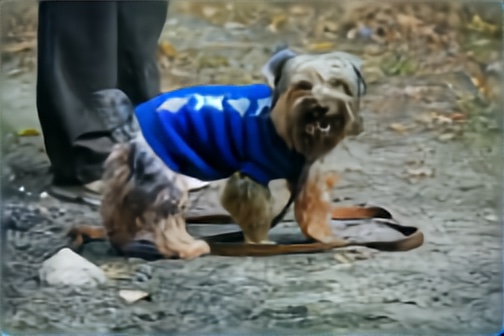}}%
        \hfill
        \subfigure[\textbf{RNN-$\frac{1}{2}$}, 0.125 bpp]{\includegraphics[width=.49\linewidth]{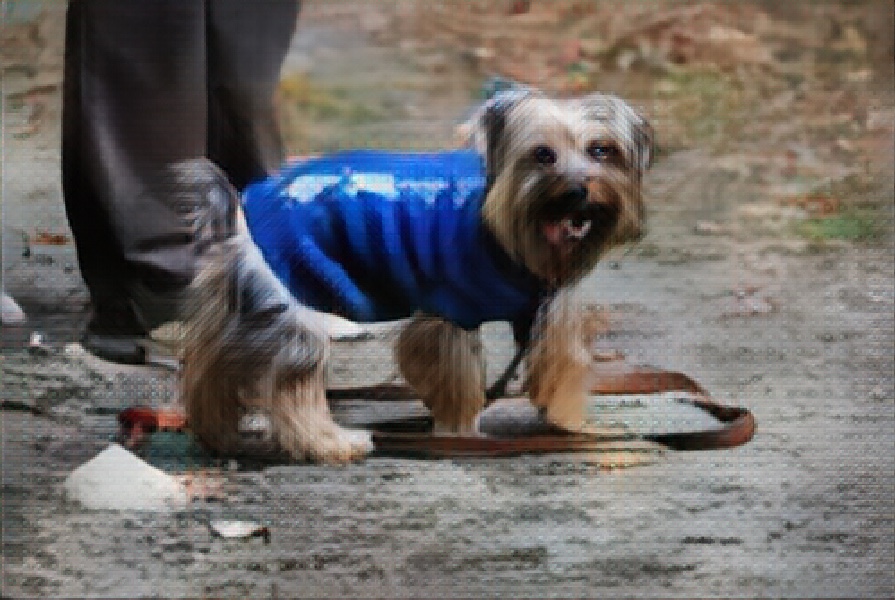}}
        \subfigure[\textbf{BPG}, 0.134 bpp]{\includegraphics[width=.49\linewidth]{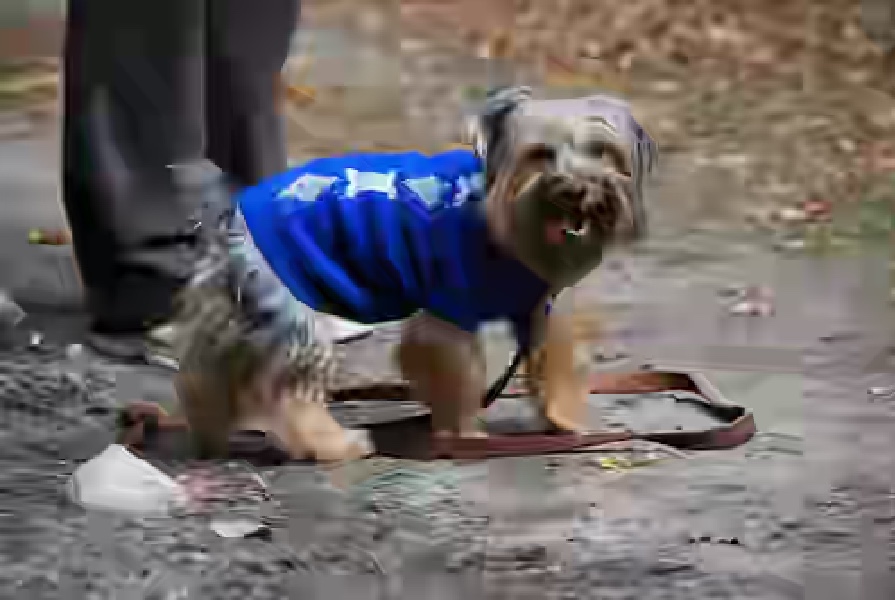}}%
        \hfill
        \subfigure[\textbf{RNN-C}, 0.125 bpp]{\includegraphics[width=.49\linewidth]{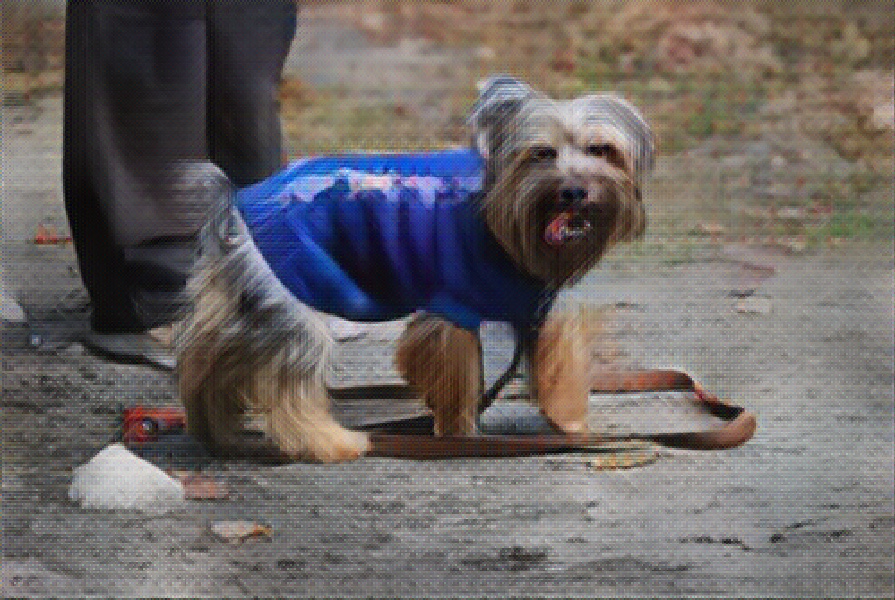}}
        \caption{Our method compared to BPG and Mentzer \etal~\cite{mentzer2018} on a sample image from Stanford Dogs.}
    \end{figure}

\begin{figure}
        \centering
        \subfigure[\textbf{Original}]{\includegraphics[width=.49\linewidth]{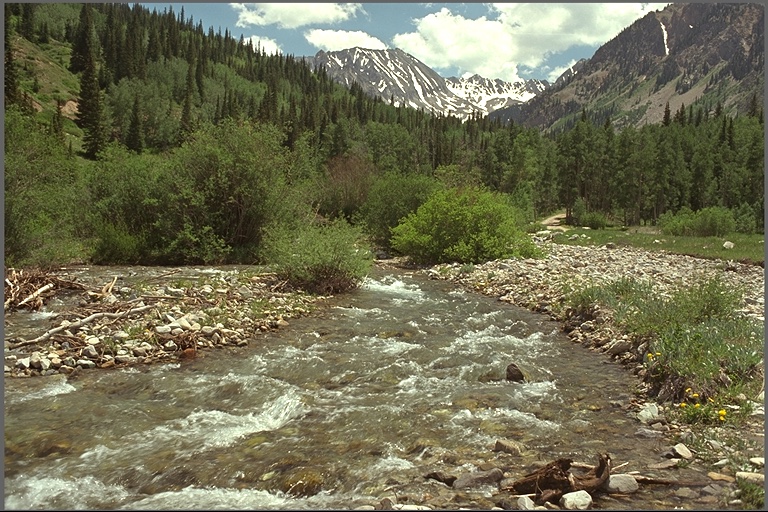}}%
        \hfill
        \subfigure[\textbf{RNN-H}, 0.125 bpp]{\includegraphics[width=.49\linewidth]{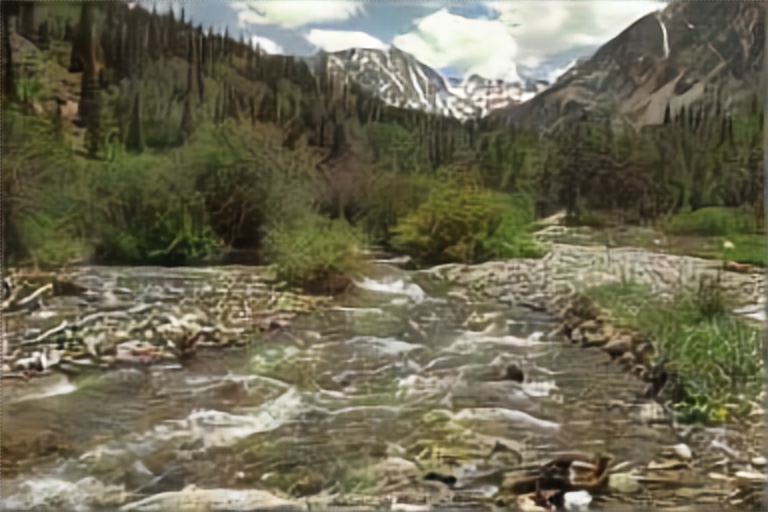}}
        \subfigure[\textbf{Mentzer \etal}~\cite{mentzer2018}, 0.195 bpp]{\includegraphics[width=.49\linewidth]{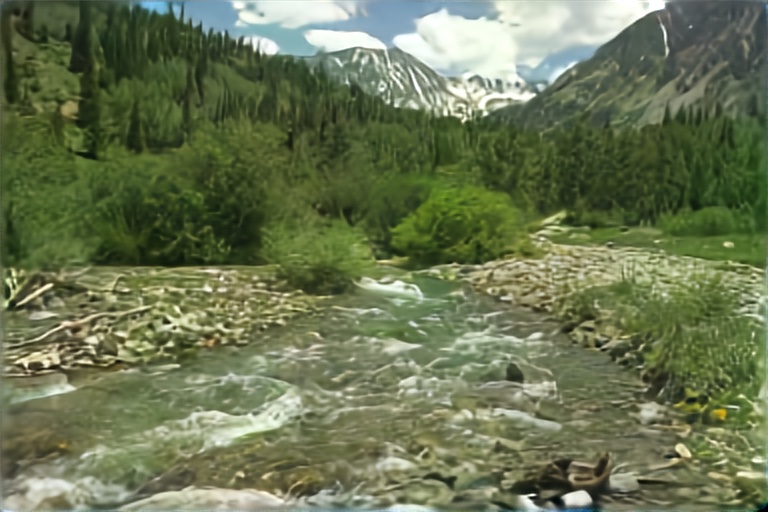}}%
        \hfill
        \subfigure[\textbf{RNN-$\frac{1}{2}$}, 0.125 bpp]{\includegraphics[width=.49\linewidth]{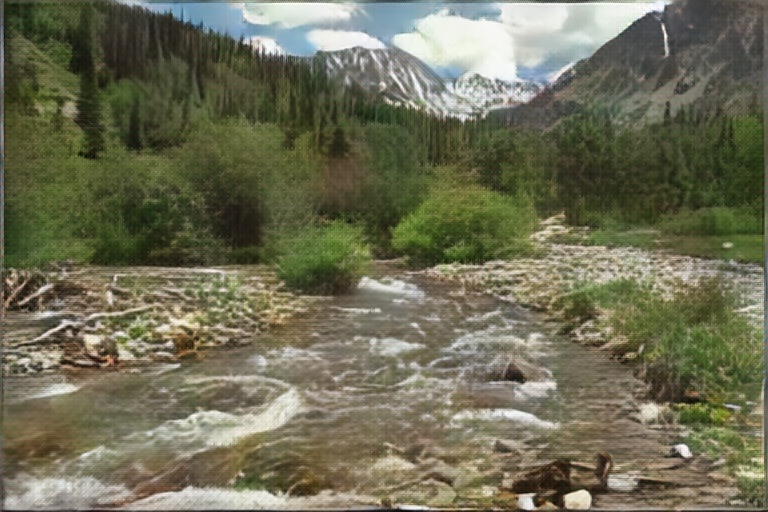}}
        \subfigure[\textbf{BPG}, 0.119 bpp]{\label{fig:kodim13_bpg}\includegraphics[width=.49\linewidth]{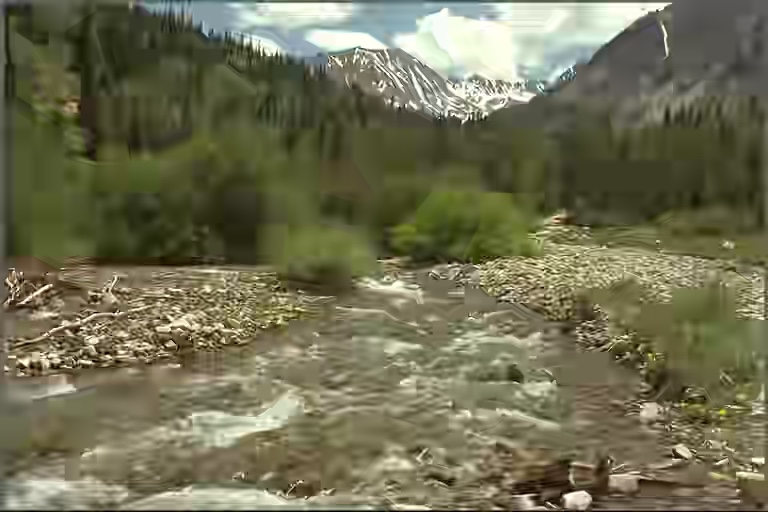}}%
        \hfill
        \subfigure[\textbf{RNN-C}, 0.125 bpp]{\includegraphics[width=.49\linewidth]{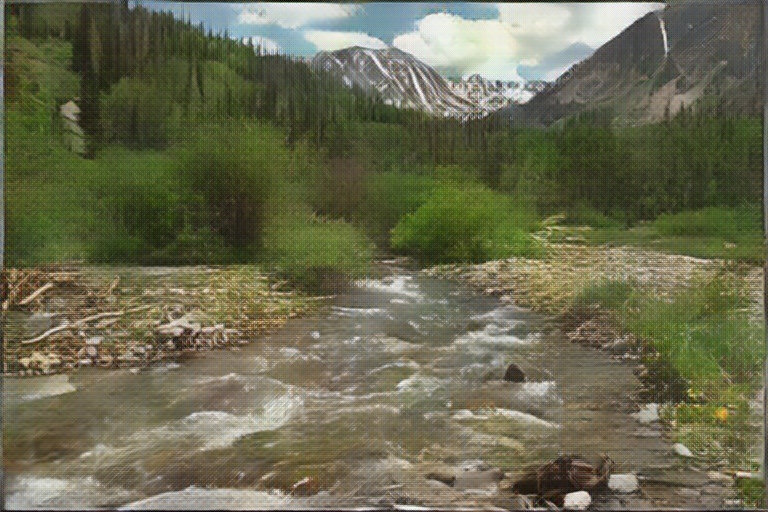}}
        \caption{Our method compared to BPG and Mentzer \etal~\cite{mentzer2018} on the 13th Kodak image.}
    \end{figure}

On the following pages, we show visual examples from each of our validation sets compressed at low bitrates. We note that all bitrates are computed without counting the header information. We generally observe that RNN-C compression exhibits high frequency artifacts noticeable to the human eye, while the MS-SSIM optimized RNN-H and Mentzer \etal~\cite{mentzer2018} results in blurring (\eg~Figures~\ref{fig:terrier_rnn_h} and~\ref{fig:terrier_cpm}). We also show each example compressed with RNN-\(\alpha\) and see that  this exhibits both types of artifacts although less noticeable. This illustrates the trade off between quality for the human observer and classification accuracy, as discussed in the main part of this paper. BPG compression typically has blocking artifacts at these compression rates as can be seen most prominently in Figure ~\ref{fig:kodim13_bpg}.
    \begin{figure}[b!]
        \centering
        \subfigure[\textbf{Original}]{\includegraphics[width=.32\linewidth]{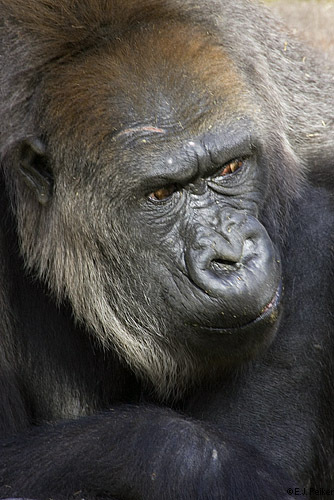}}%
        \hfill
        \subfigure[\textbf{Mentzer \etal}~\cite{mentzer2018}, 0.153 bpp]{\includegraphics[width=.32\linewidth]{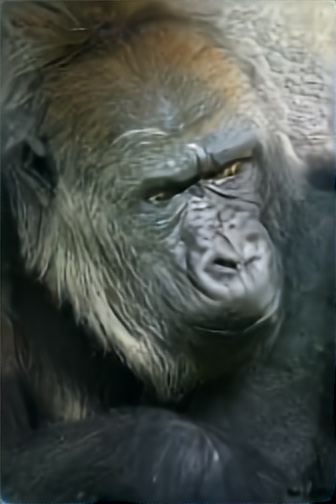}}%
        \hfill
        \subfigure[\textbf{BPG}, 0.131 bpp]{\includegraphics[width=.32\linewidth]{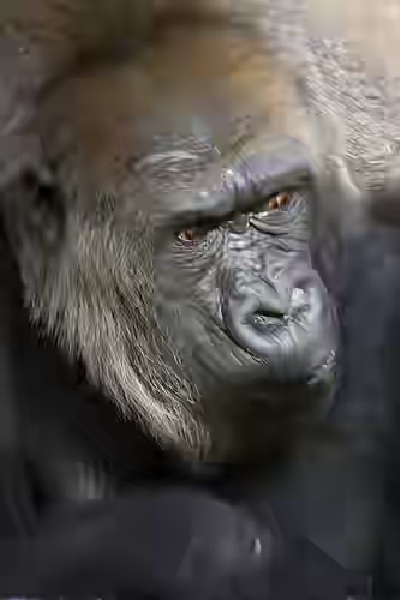}}
        \subfigure[\textbf{RNN-H}, 0.125 bpp]{\label{fig:gorilla_rnn_h} \includegraphics[width=.32\linewidth]{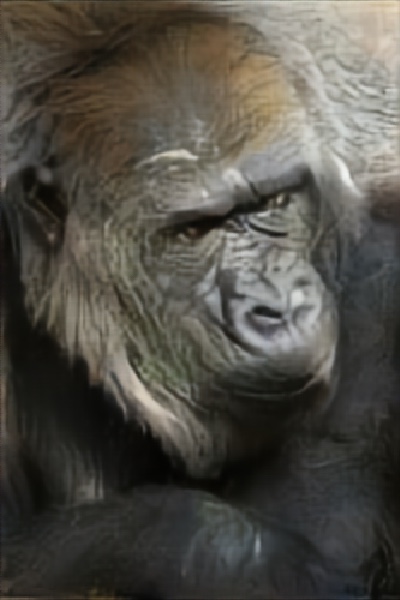}}%
        \hfill
        \subfigure[\textbf{RNN-$\frac{1}{2}$}, 0.125 bpp]{\includegraphics[width=.32\linewidth]{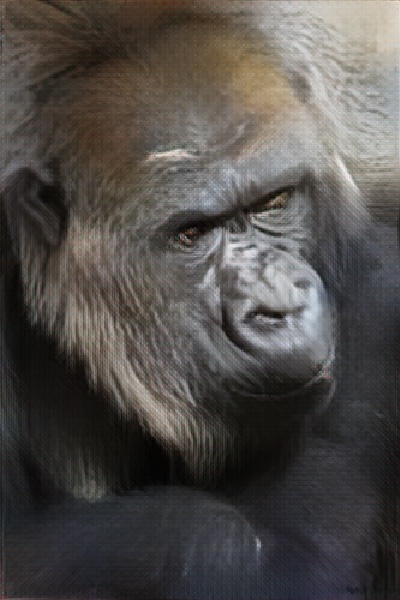}}%
        \hfill
        \subfigure[\textbf{RNN-C}, 0.125 bpp]{\includegraphics[width=.32\linewidth]{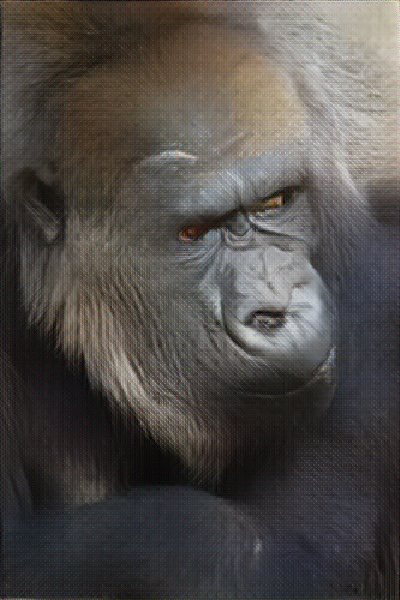}}
        \caption{Our method compared to BPG and Mentzer \etal~\cite{mentzer2018} on a sample image from ImageNet-1K.}
    \end{figure}
    \begin{figure}
        \centering
        \subfigure[\textbf{Original}]{\includegraphics[width=.45\linewidth]{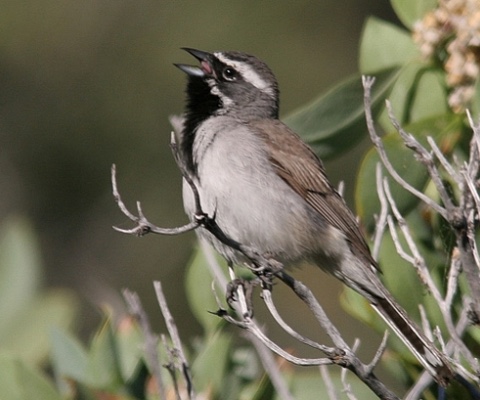}}%
        \hfill
        \subfigure[\textbf{RNN-H}, 0.125 bpp]{\includegraphics[width=.45\linewidth]{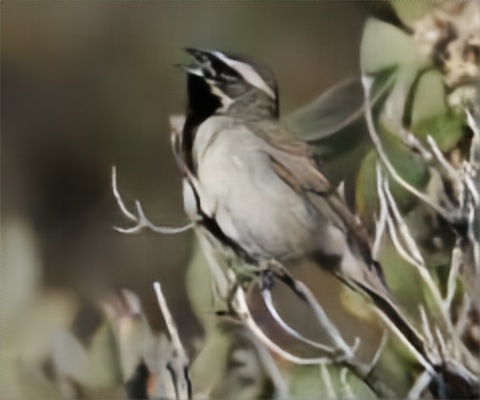}}
        \subfigure[\textbf{Mentzer \etal}~\cite{mentzer2018}, 0.119 bpp]{\includegraphics[width=.45\linewidth]{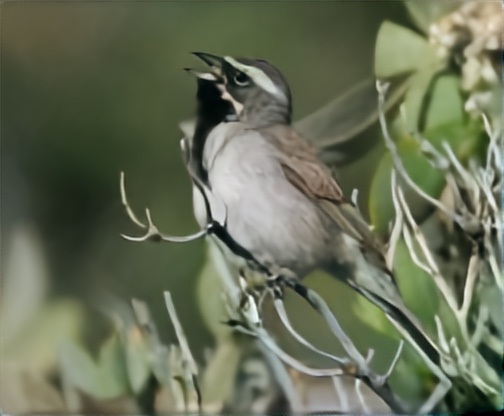}}%
        \hfill
        \subfigure[\textbf{RNN-$\frac{1}{2}$}, 0.125 bpp]{\includegraphics[width=.45\linewidth]{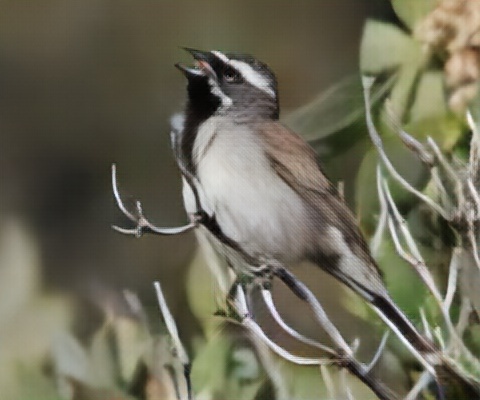}}
        \subfigure[\textbf{BPG}, 0.118 bpp]{\includegraphics[width=.45\linewidth]{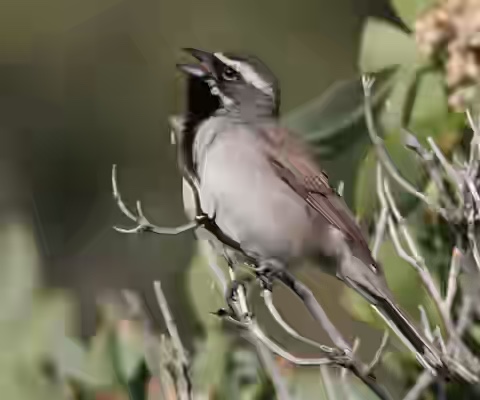}}%
        \hfill
        \subfigure[\textbf{RNN-C}, 0.125 bpp]{\includegraphics[width=.45\linewidth]{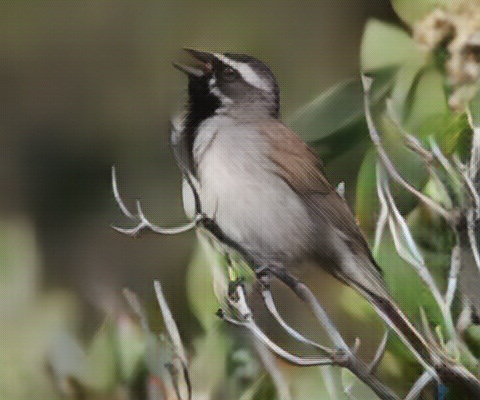}}
        \caption{Our method compared to BPG and Mentzer \etal~\cite{mentzer2018} on a sample image from CUB-200-2011.}
    \end{figure}
    \begin{figure}
        \centering
        \subfigure[\textbf{Original}]{\includegraphics[width=.49\linewidth]{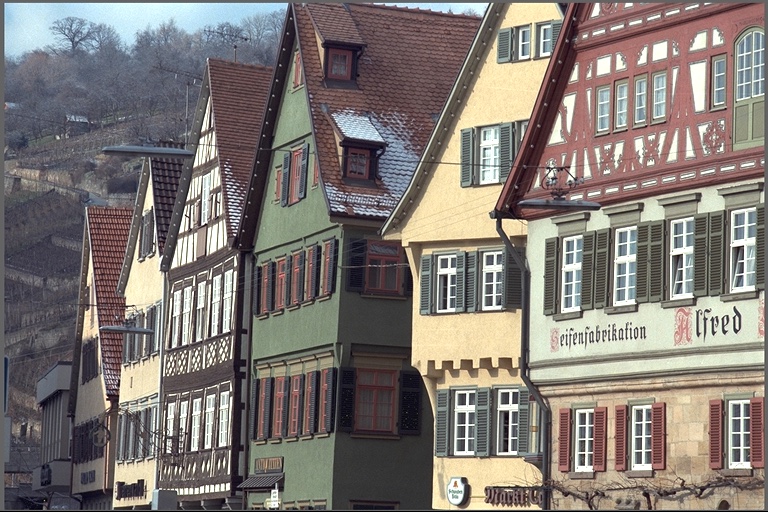}}%
        \hfill
        \subfigure[\textbf{RNN-H}, 0.125 bpp]{\includegraphics[width=.49\linewidth]{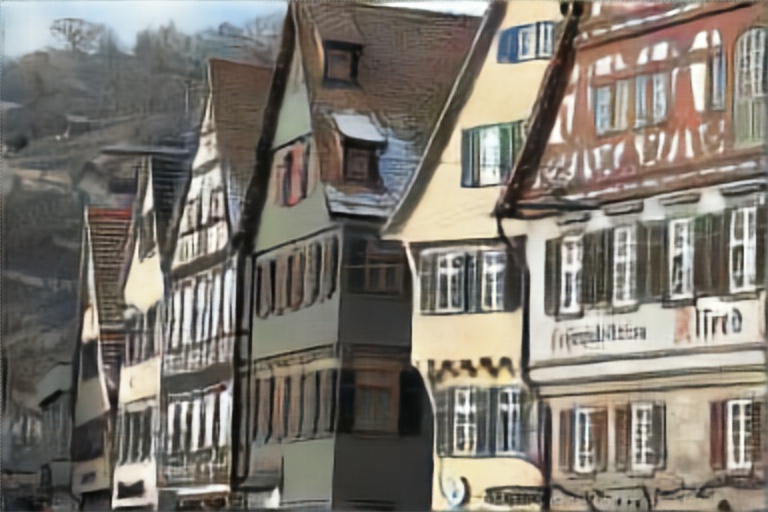}}
        \subfigure[\textbf{Mentzer \etal}~\cite{mentzer2018}, 0.175 bpp]{\includegraphics[width=.49\linewidth]{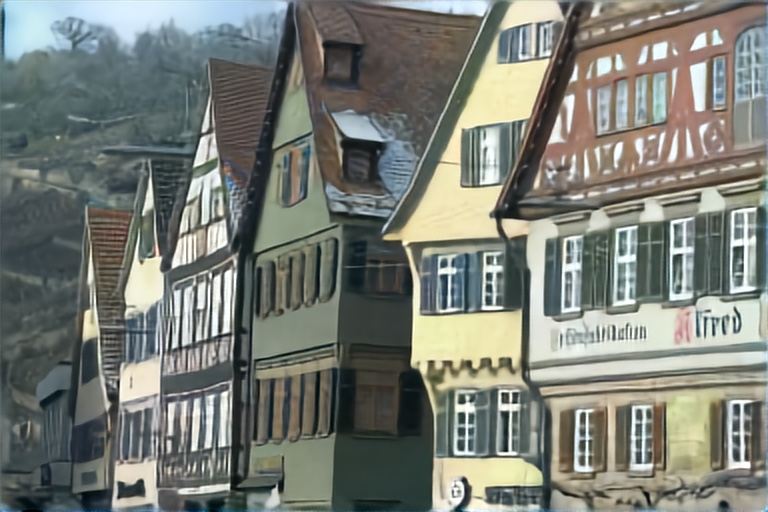}}%
        \hfill
        \subfigure[\textbf{RNN-$\frac{1}{2}$}, 0.125 bpp]{\includegraphics[width=.49\linewidth]{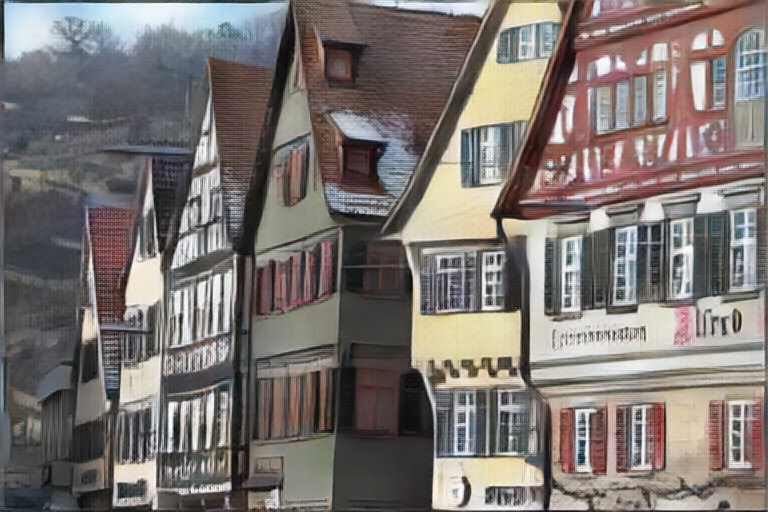}}
        \subfigure[\textbf{BPG}, 0.141 bpp]{\includegraphics[width=.49\linewidth]{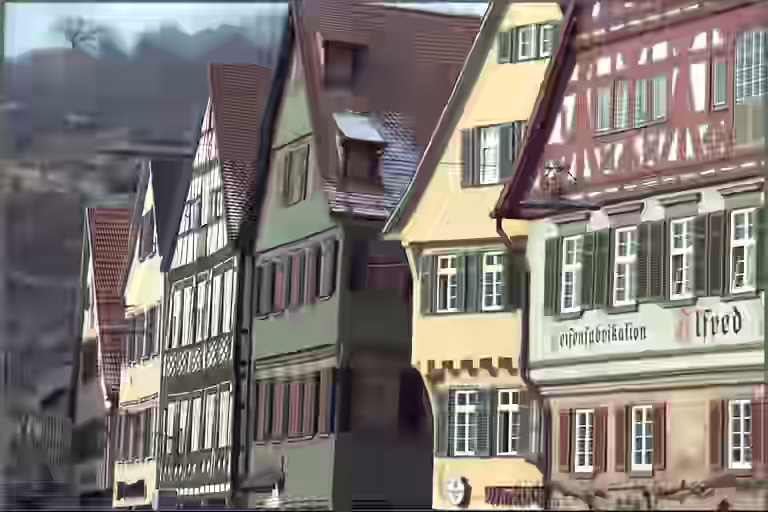}}%
        \hfill
        \subfigure[\textbf{RNN-C}, 0.125 bpp]{\includegraphics[width=.49\linewidth]{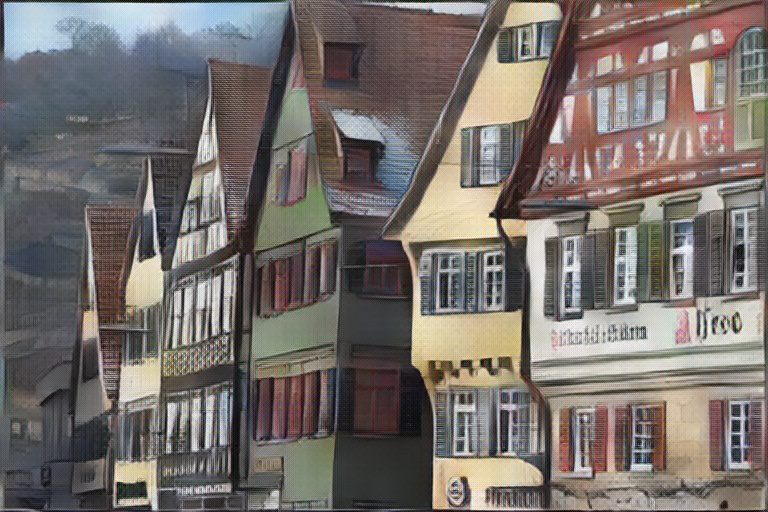}}
        \caption{Our method compared to BPG and Mentzer \etal~\cite{mentzer2018} on the 8th Kodak image.}
    \end{figure}
    \begin{figure}
        \centering
        \subfigure[\textbf{Original}]{\includegraphics[width=.32\linewidth]{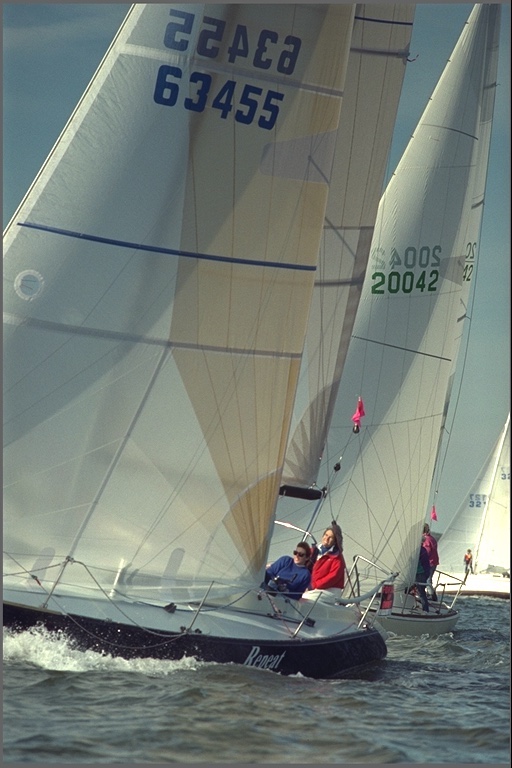}}%
        \hfill
        \subfigure[\textbf{Mentzer \etal}~\cite{mentzer2018}, 0.125 bpp]{\includegraphics[width=.32\linewidth]{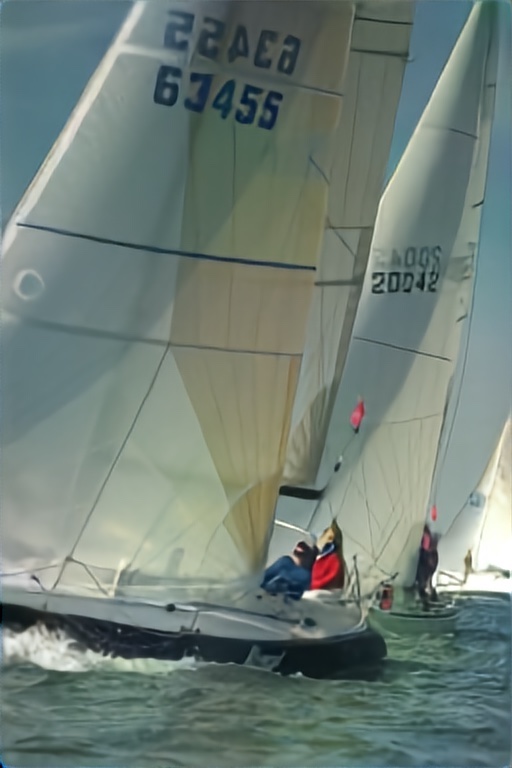}}%
        \hfill
        \subfigure[\textbf{BPG}, 0.131 bpp]{\includegraphics[width=.32\linewidth]{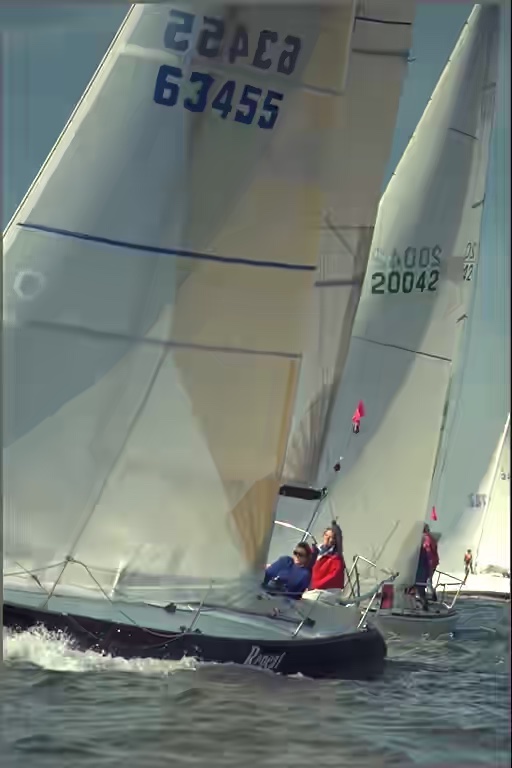}}
        \subfigure[\textbf{RNN-H}, 0.125 bpp]{\includegraphics[width=.32\linewidth]{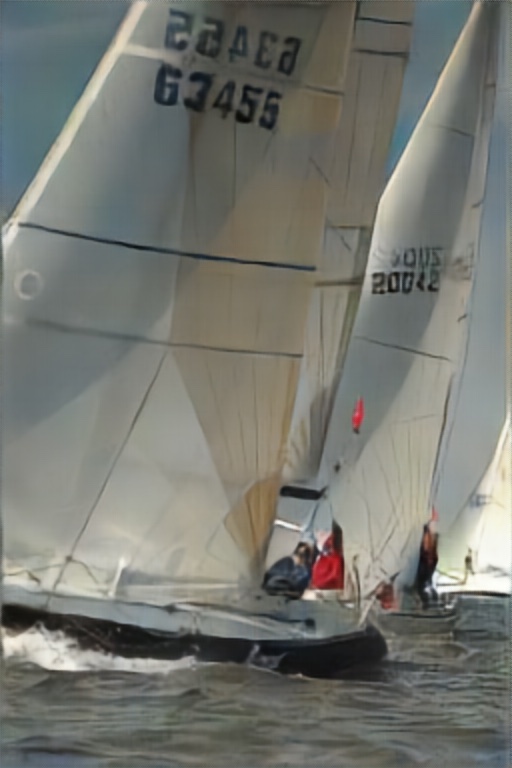}}%
        \hfill
        \subfigure[\textbf{RNN-$\frac{1}{2}$}, 0.125 bpp]{\includegraphics[width=.32\linewidth]{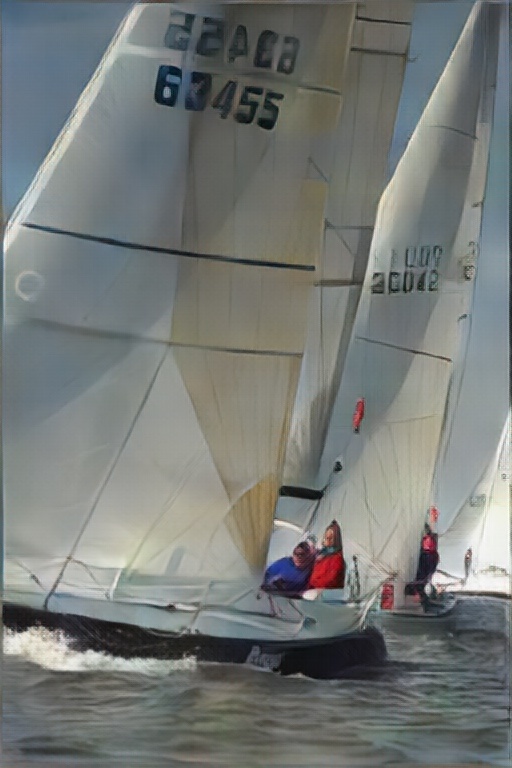}}%
        \hfill
        \subfigure[\textbf{RNN-C}, 0.125 bpp]{\includegraphics[width=.32\linewidth]{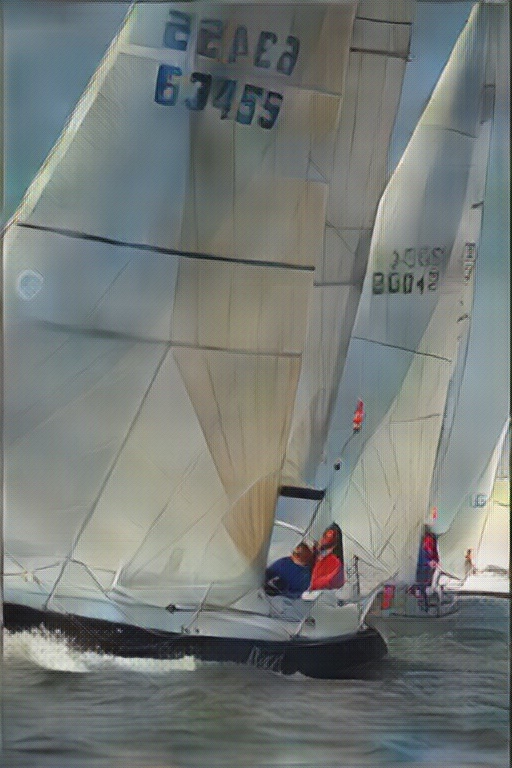}}
        \caption{Our method compared to BPG and Mentzer \etal~\cite{mentzer2018} on the 10th Kodak image.}
    \end{figure}
